\documentstyle[aps,eqsecnum,preprint,floats,epsf]{revtex}

\begin{document}

\tighten
\draft

\preprint{
\begin{minipage}{3.5cm}
Budker INP 96-45\\
CERN-TH/96-144\\
hep-ph/9607366\\
\end{minipage}
}

\title{Asymptotics of heavy-meson form factors}

\author{A.G.~Grozin}
\address{Budker Institute of Nuclear Physics, Novosibirsk 630090, 
Russia}

\author{M.~Neubert}
\address{Theory Division, CERN, CH--1211 Geneva 23, Switzerland}


\maketitle

\begin{abstract}
Using methods developed for hard exclusive QCD processes, we
calculate the asymptotic behaviour of heavy-meson form factors at
large recoil. It is determined by the leading- and subleading-twist
meson wave functions. For $1\ll |v\cdot v'|\ll m_Q/\Lambda$, the form
factors are dominated by the Isgur--Wise function, which is
determined by the interference between the wave functions of leading
and subleading twist. At $|v\cdot v'|\gg m_Q/\Lambda$, they are
dominated by two functions arising at order $1/m_Q$ in the
heavy-quark expansion, which are determined by the leading-twist wave
function alone. The sum of these contributions describes the form
factors in the whole region $|v\cdot v'|\gg 1$. As a consequence,
there is an exact zero in the form factor for the scattering of
longitudinally polarized $B^*$ mesons at some value $v\cdot v'\sim
m_b/\Lambda$, and an approximate zero in the form factor of $B$
mesons in the timelike region ($v\cdot v'\sim -m_b/\Lambda$). We
obtain the evolution equations and sum rules for the wave functions
of leading and subleading twist as well as for their moments. We
briefly discuss applications to heavy-meson pair production in
$e^+e^-$ collisions.
\end{abstract}


\narrowtext

\section{Introduction}
\label{Intro}

In the heavy-quark effective theory
(HQET)~\cite{IW,EiHi,Grin,Geor,FGL}
(see~\cite{GeRev,IWRev,GrRev,review,MaRev,myrev} for reviews), the
heavy-quark spin does not interact with gluons to leading order in
$1/m_Q$ (where $m_Q$ is the heavy-quark mass). Therefore, this spin
can be rotated (spin symmetry) or even switched off (superflavour
symmetry~\cite{Super,Caro}) without affecting the dynamics. In the
heavy-quark limit, the properties of the doublet of the ground-state
pseudoscalar and vector mesons $(q\bar Q)$ are therefore
characterized by the spin-parity quantum numbers $j^P=\frac12^+$ of
the light degrees of freedom~\cite{IW,AFal}. In this paper, we shall
use the superflavour symmetry to describe the ground-state mesons by
a Dirac wave function. However, we collect in Appendix~A the most
important formulae using a more conventional formalism.

Let $Q_v^*$ be a scalar field describing a heavy antiquark moving at
four-velocity $v$ with its spin switched off. Then the decay constant 
$f$ of a heavy meson (moving at the same velocity) is defined as
\begin{equation}
   \langle\,0\,|Q_v^* q|M(v)\rangle = f\,u(v) \,,
\label{i1}
\end{equation}
where $u(v)$ is the Dirac wave function of the meson, which satisfies
\begin{equation}
   \rlap/v\,u(v) = u(v) \,.
\end{equation}
A non-relativistic (i.e.\ mass independent) normalization of $u(v)$
and $|M(v)\rangle$ is assumed. In the heavy-quark limit, the relation
between $f$ and the usual meson decay constants reads
\begin{equation}
   f_M = f_{M^*} = \frac{2 f}{\sqrt{m_Q}} \,.
\label{i2}
\end{equation}

To leading order in $1/m_Q$, current-induced transitions between two
ground-state mesons are described by a single Isgur-Wise form
factor~\cite{IW,Falk}:
\begin{equation}
   \langle M(v')|Q_v^* Q_{v'}|M(v)\rangle
   = \xi(v\cdot v')\,\bar u(v')\,u(v) \,,
\label{i3}
\end{equation}
where $v$, $v'$ are the meson velocities. At next-to-leading order,
there appear $1/m_Q$ corrections to the currents and to the
Lagrangian of the HQET~\cite{Luke}. The first type of corrections can
be expressed via the matrix element of a dimension-four
operator:\footnote{We use $D^\mu=\partial^\mu-i A^\mu$ and
$D^{\mu\dagger}=\partial^\mu+i A^{\mu\dagger}$, where $A^\mu=g_s t_a
A_a^\mu$ is the gluon field.}
\begin{equation}
   \langle M(v')|(i D^{\mu\dagger} Q_v^*) Q_{v'}|M(v)\rangle
   = \bar u(v')\,\xi^\mu(v,v')\,u(v) \,,
\label{i4a}
\end{equation}
where
\begin{equation}
   \xi^\mu(v,v') = \xi_+(v\cdot v') (v+v')^\mu
   + \xi_-(v\cdot v') (v-v')^\mu 
   + \xi_3(v\cdot v')\,\gamma^\mu \,.
\label{i4}
\end{equation}
The equations of motion, $iv\cdot D Q_v=0$, can be employed to relate
$\xi_\pm$ to $\xi_3$~\cite{Luke}. The result is:
\begin{equation}
   \xi_+ = \frac{\case 12 (v\cdot v'-1)\,\bar\Lambda\,\xi - \xi_3}
   {v\cdot v'+1} \,,\quad
   \xi_- = \case 12\,\bar\Lambda\,\xi \,,
\label{i6}
\end{equation}
where $\bar\Lambda$ is the ``binding energy'', i.e.\ the difference
between the meson mass and the heavy-quark mass. Hence, there is only
one new independent form factor. The $1/m_Q$ corrections to the
Lagrangian give rise to the matrix elements of two non-local
operators:
\begin{eqnarray}
   \langle M(v')|\,i\!\int\!\text{d}x\,
   T\{ Q_v^* Q_{v'}(0), Q_v^* (iD)^2 Q_v(x) \}|M(v)\rangle
   &=& 2\chi_1(v\cdot v')\,\bar u(v') u(v) \,, \nonumber\\
   \langle M(v')|\,i\!\int\!\text{d}x\,
   T\{ Q_v^* Q_{v'}(0), Q_v^*\,i G^{\mu\nu} Q_v(x) \}
   |M(v)\rangle &=& 2\bar u(v')\,\chi^{\mu\nu}(v,v')\,u(v) \,,
\label{chidef}
\end{eqnarray}
where $iG^{\mu\nu}=[iD^\mu,iD^\nu]=i g_s t_a G_a^{\mu\nu}$ is the
gluon field strength, and ($\sigma^{\mu\nu}=\frac{i}{2}[\gamma^\mu,
\gamma^\nu]$)
\begin{equation}
   \chi^{\mu\nu}(v,v') = \chi_2(v\cdot v')(\gamma^\mu v^{\prime\nu} 
   - \gamma^\nu v^{\prime\mu}) + 2i \chi_3(v\cdot v')
   \sigma^{\mu\nu} \,.
\label{i7}
\end{equation}
In the second matrix element in (\ref{chidef}), the indices $\mu$ and
$\nu$ are restricted to the subspace orthogonal to $v$. Explicit
expressions for the meson form factors in terms of the functions
$\xi$, $\xi_3$ and $\chi_i$ can be found in~\cite{review,Luke}. At
moderate values of $v\cdot v'$, these functions have been studied
extensively in the framework of QCD sum rules, both at
leading~\cite{Rad,N1,BS,BBG,N2} and next-to-leading
\cite{N3,BaiG,NLN} order in the heavy-quark expansion. In the present
paper, we shall consider the behaviour of the form factors in the
large-recoil region $|v\cdot v'|\gg 1$. This region is inaccessible
in the weak semileptonic decays, but it can be explored (at least in
principle) in the production reaction $e^+e^-\to M^{(*)}\bar
M^{(*)}$. Using methods developed for hard exclusive processes, we
calculate the asymptotic behaviour of the form factors in a
model-independent way.

\begin{figure}
\epsfxsize=7cm
\centerline{\epsffile{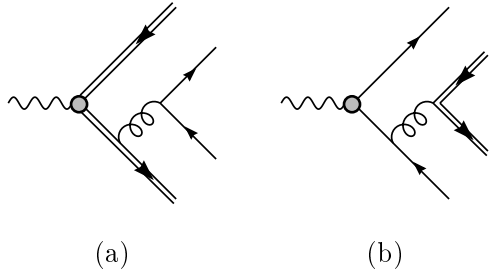}}
\vspace{0.2cm}
\caption{Hard-gluon exchange contributions to heavy-meson form
factors. The external current is presented by the wave line; the
heavy antiquark is represented by a double line.}
\label{ffdiag}
\end{figure}

Our results can be summarized as follows: For $|v\cdot v'|\gg 1$,
there is a large momentum transfer to the light quark:
$q_{\text{light}}^2\sim -\Lambda^2\,v\cdot v'$, where $\Lambda$ is of
the order of a typical hadronic mass scale. As shown in
Fig.~\ref{ffdiag}a, this momentum is transferred by the exchange of a
hard gluon, and the methods developed for hard exclusive processes in
QCD~\cite{Excl1,Excl2,Excl3,Brodsky} (see~\cite{Chernyak,Baier,Brev}
for reviews) are applicable. In the ``brick wall'' frame, where
$\vec{v}\,'=-\vec{v}$, the projection of the total angular momentum
on the $z$ axis (directed along $\vec{v}$) is equal to the projection
of the meson spin. (Recall that the heavy-quark spin has been
switched off.) Since it is conserved, the meson helicity changes its
sign~\cite{Politzer}. The asymptotic behaviour of the Isgur--Wise
form factor is thus determined by the interference between the
leading-twist (quark--antiquark) wave function and the
subleading-twist (quark--antiquark and quark--antiquark--gluon) wave
functions. It is given by
\begin{equation}
   \xi(v\cdot v') \sim \frac{\alpha_s f^2}{\Lambda^3 (v\cdot v')^2} \,.
\label{i8}
\end{equation}
Indeed, the situation is similar to the well-known case of the
$\pi$--$\rho$ form factor~\cite{Chernyak}. At order $1/m_Q$ in the
heavy-quark expansion, there are however contributions involving the
leading-twist wave function only. They conserve the meson helicity
and behave as
\begin{equation}
   \frac{\xi_3(v\cdot v')}{m_Q} \sim
   \frac{v\cdot v'\,\chi_2(v\cdot v')}{m_Q} \sim
   \frac{\alpha_s f^2}{m_Q\Lambda^2\,v\cdot v'} \,.
\label{i9}
\end{equation}
The leading contribution in the heavy-quark expansion, which is given
by the Isgur--Wise function, dominates as long as $|v\cdot v'|\ll
m_Q/\Lambda$. For $|v\cdot v'|\gg m_Q/\Lambda$, however, the
contributions of $\xi_3$ and $\chi_2$ in (\ref{i9}) become the
dominant ones. Note that they violate the heavy-quark spin symmetry,
i.e.\ they contribute in a non-universal way to the various meson
form factors. Higher-order terms in the heavy-quark expansion (of
order $1/m_Q^2$ and higher) cannot fall off slower than $1/(v\cdot
v')$ because this behaviour corresponds to the leading twist, and
hence they always remain small corrections.

It is instructive to consider the same situation from an opposite
point of view. At asymptotically large values $|q^2|\simeq 2
m_Q^2|v\cdot v'|\gg m_Q^2/x$ (where $q$ is the momentum transferred
to the mesons, and $x\sim\Lambda/m_Q$ is the momentum fraction
carried by the light quark), the form factor of a heavy meson behaves
like that of the
pion~\cite{Excl1,Excl2,Excl3,Brodsky,Chernyak,Baier,Brev}:
\begin{equation}
   F(q^2) \sim \frac{\alpha_s f_M^2}{x^2 q^2} \,,
\label{i10}
\end{equation}
which exactly corresponds to~(\ref{i9}). However, there is a
contribution to the subleading-twist ($1/q^4$) correction which is
proportional to $m_Q^2$. It becomes important for moderate values of
$q^2$:
\begin{equation}
   F(q^2) \sim \frac{\alpha_s m_Q^2 f_M^2}{x^3 q^4} \,.
\label{i11}
\end{equation}
This contribution exactly corresponds to~(\ref{i8}). It dominates for
$|q^2|\ll m_Q^2 x$. Higher-twist ($1/q^6$ and higher) corrections
cannot be more enhanced than $m_Q^2$ because otherwise the form
factor would diverge in the heavy-quark limit, and hence they always
remain small corrections.

Until now, we considered form factors for transitions induced by a
current containing heavy quarks only. In the case of, say, the
electromagnetic current, which also contains light-quark fields,
contributions of the type shown in Fig.~\ref{ffdiag}b appear.
However, they lead to the behaviour
\begin{equation}
   \frac{\alpha_s f^2}{m_Q^3\,v\cdot v'} \sim
   \frac{\alpha_s f_M^2}{q^2}
\label{i12}
\end{equation}
and can thus be safely neglected, since $m_Q\gg\Lambda$ for a heavy
quark.

In summary, for $1\ll|v\cdot v'|\ll m_Q/\Lambda$ the dominant
contribution to meson form factors comes from the universal
Isgur--Wise function. It involves the exchange of a longitudinal hard
gluon and a change of the meson helicity, corresponding to the
interference between the leading- and subleading-twist wave
functions. For $|v\cdot v'|\gg m_Q/\Lambda$, on the other hand, the
situation is the same as in massless QCD. The asymptotic behaviour of
meson form factors is determined by leading-twist contributions,
which are not universal. They are governed by the exchange of a hard
transverse gluon, which conserves the meson helicity. The sum of
these two contributions describes the asymptotic behaviour of the
form factors in the whole region $|v\cdot v'|\gg 1$. The simple
picture described here is only slightly modified by the emission of
gluon bremsstrahlung, which can be dealt with in
renormalization-group improved perturbation theory. It leads to an
additional, moderate power suppression of the form factors at large
recoil.

The remainder of the paper is organized as follows: In Sec.~\ref{WF},
we introduce the (quark--antiquark) meson wave functions of leading
and subleading twist, $\varphi_\pm(\omega)$. Unlike in QCD, wave
functions defined in the HQET depend on a dimensional argument
$\omega$. We investigate the moments of these wave functions and
derive the symmetry relations between the various meson wave
functions, which arise in the heavy-quark limit. In Sec.~\ref{Evol},
we derive the evolution equations for $\varphi_\pm(\omega)$, which
are analogous to the Brodsky--Lepage equations~\cite{Brodsky}. This
calculation extends the calculation of the HQET anomalous dimension
of local heavy-light current
operators~\cite{OneLoop,Vol1,PoWi,TwoLoop,BrGr,Gime}. A new kind of
ultraviolet divergence appears in the relation between the local
operators and the operators defining the wave function. Therefore,
the Brodsky--Lepage kernels do not determine the renormalization
properties of the local operators completely. A similar situation is
encountered in the case of the (Altarelli--Parisi) equations
describing the evolution of distribution functions in the
HQET~\cite{Grisha}. In Sec.~\ref{SR}, we investigate the properties
of the wave functions $\varphi_\pm(\omega)$ using the QCD sum-rule
approach. This extends the heavy-meson sum
rules~\cite{N1,Patricia,David,Shuryak} to the case of non-local
operators. After considering the sum rules for the lowest moments, we
construct the sum rules for the wave functions themselves, taking
into account the non-locality of the quark condensate~\cite{Mikh}. In
Sec.~\ref{IW2}, we apply our results to derive the leading asymptotic
behaviour of meson form factors at large recoil. First, we calculate
the contribution of the interference between the leading-twist and
the (quark--antiquark) subleading-twist wave functions to the
asymptotic behaviour of the Isgur--Wise function.\footnote{The
properties of quark--antiquark--gluon subleading-twist wave functions
and their contribution to $\xi(v\cdot v')$ will be discussed
elsewhere.}
Then we calculate the leading-twist contributions to the form factors
appearing at order $1/m_Q$ in the heavy-quark expansion. Finally, in
Sec.~\ref{apps} we discuss the implications of our results for the
reactions $e^+e^-\to B^{(*)}\bar B^{(*)}$ and $e^+e^-\to D^{(*)}\bar
D^{(*)}$. Technical details of our calculations are presented in four
appendices.

Before we proceed, some comments on the existing literature on the
application of perturbative QCD to the calculation of heavy-meson
form factors are in order. A simple model of a meson made out of two
heavy quarks with unequal masses was considered in~\cite{zero}. There
it was noted that some form factors must have zeros in the physical
region. We confirm this interesting observation, although we disagree
with some other results of this work (see Sec.~\ref{IW2}). A similar
model was considered in~\cite{KK,Seoul,Milana}. There, a single spin
structure of the heavy-meson wave function was used, which is
determined from the condition that the light quark be at rest in the
meson rest frame. Hence, all quark--antiquark wave functions have the
same shape, and this shape was assumed to be $\delta(\omega-\mu)$,
with $\mu$ being the constituent mass of the light quark. As we shall
see later, for a realistic heavy meson the wave functions
$\varphi_+(\omega)$ and $\varphi_-(\omega)$ do not coincide, and they
are not well approximated by sharply peaked functions. Integrals for
the form factors receive important contributions from the region of
low $\omega$ values, which are missing in the peaking approximation.
Therefore, the results obtained using such a static quark model can
at best be taken as a crude estimate. Perturbative QCD and the
constituent quark model were recently applied also to the
semileptonic decays $\bar B\to D^{(*)}\ell\,\bar\nu$~\cite{Li}, for
which $1<v\cdot v'<1.6$. In our opinion, such small values of $v\cdot
v'$ are far too low to treat the gluon with $k_g^2\sim
-\Lambda^2\,v\cdot v'$ perturbatively. Moreover, the calculations
in~\cite{Li} are done using a model wave function with an ad hoc
$k_\bot$ dependence, whose longitudinal-momentum dependence
contradicts the expectations based on the HQET.

\section{Quark--Antiquark Wave Functions}
\label{WF}

We shall define the quark--antiquark wave functions
$\widetilde\varphi_\pm(t)$ of a heavy meson in terms of the matrix
element of the bilocal operator
\begin{equation}
   \widetilde O(t) = Q^*(0)\,E(0,z)\,q(z) \,;\quad t=v\cdot z \,,
\label{wf1}
\end{equation}
where $z$ is a null vector on the light cone ($z^2=0$), and
\begin{equation}
   E(x,y) = P\exp\Bigg( -i\int\limits_x^y\!\text{d}z^\mu\,A_\mu(z)
   \Bigg)
\label{string}
\end{equation}
is a string operator ensuring gauge invariance. In the light-cone
gauge ($A_+=0$), one simply has $E(0,z)=1$. Since in this section we
are considering operators containing a single heavy quark field
$Q_v^*$, we shall for simplicity omit the velocity label on the
field. Similarly, we shall write $M$ and $u$ instead of $M(v)$ and
$u(v)$. The meson matrix element of the operator $\widetilde O(t)$
has two independent Dirac structures, $u$ and $\rlap/z\,u$, and we
define
\begin{equation}
   \langle\,0\,|\widetilde O(t)|M\rangle
   = f\,\Big( \widetilde\varphi_+(t) + \frac{1}{2t}\,
   [\widetilde\varphi_-(t) - \widetilde\varphi_+(t)]\,\rlap/z
   \Big)\,u \,.
\label{wf2}
\end{equation}

It is convenient to introduce two light-cone vectors
$n_\pm^\mu=(1,0,0,\mp 1)$ such that $n_\pm^2=0$ and $n_+\cdot n_-=2$.
Any vector $a^\mu$ can be decomposed as $a^\mu=\frac 12(a_+ n_-^\mu +
a_- n_+^\mu) + a_\bot^\mu$, where $a_\pm=a\cdot n_\pm$. This implies
$a\cdot b=\frac 12(a_+ b_- + a_- b_+) - \vec a_\bot\cdot\vec b_\bot$.
We shall also use the light-cone components of the Dirac matrices,
defined as $\gamma_\pm = n_\pm^\mu\gamma_\mu = \rlap/n_\pm$. If the
meson is at rest, then $v^\mu=\case 12 (n_+^\mu + n_-^\mu)$, i.e.\
$v_+=v_-=1$. Using $\rlap/v\,u=u$, we can then rewrite (\ref{wf2}) as
\begin{equation}
   \langle\,0\,|\widetilde O(t)|M\rangle
   = \case 12 f\,\Big[ \widetilde\varphi_+(t) \gamma_-
   + \widetilde\varphi_-(t) \gamma_+ \Big]\,u \,.
\label{wf2a}
\end{equation}
For a meson with an arbitrary velocity in the $n_+$--$n_-$ plane,
this formula becomes
\begin{equation}
   \langle\,0\,|\widetilde O(t)|M\rangle
   = \case 12 f\,\Big[ \widetilde\varphi_+(t)\,v_+ \gamma_-
   + \widetilde\varphi_-(t)\,v_- \gamma_+ \Big]\,u \,.
\label{wf2b}
\end{equation}
If we introduce the rapidity $\vartheta$ by writing 
\begin{equation}
   v^\mu = (\cosh\vartheta,0,0,\sinh\vartheta) \,,
\end{equation}
then $v_+=e^\vartheta$ and $v_-=e^{-\vartheta}$. This shows that for
a fast-moving meson ($\vartheta\gg 0$), $\widetilde\varphi_+$ is the
leading-twist wave function, whereas $\widetilde\varphi_-$ has
subleading twist. It is convenient to project onto these wave
functions by writing
\begin{equation}
   \langle\,0\,|\widetilde O_\pm(t)|M\rangle
   = f\,\widetilde\varphi_\pm(t)\,\gamma_\pm u \,,\quad
   \widetilde O_\pm(t)=\gamma_\pm \widetilde O(t) \,.
\label{Opmt}
\end{equation}
This result is valid in an arbitrary reference frame, as can be seen
by using the relations
\begin{equation}
   \gamma_\pm^2 = 0 \,,\quad \gamma_\pm\gamma_\mp
   = \frac{2}{v_\pm}\,\gamma_\pm\rlap/v \,.
\label{gampm}
\end{equation}

The wave functions $\widetilde\varphi_\pm(t)$ depend on the
separation $t$ on the light cone. We define the corresponding wave
functions in momentum space by
\begin{equation}
   \varphi_\pm(\omega) = \frac{1}{2\pi} \int\text{d}t\,
   \widetilde\varphi_\pm(t)\,e^{i\omega t} \,,\quad
   \widetilde\varphi_\pm(t) = \int\text{d}\omega\,
   \varphi_\pm(\omega)\,e^{-i\omega t} \,.
\label{wf4}
\end{equation}
The variable $\omega$ has the meaning of the light-cone projection
$p_+$ of the light-quark momentum in the heavy-meson rest frame. The
positions of the singularities in the complex $t$ plane are such that
$\varphi_\pm(\omega)$ vanish for $\omega<0$. The wave functions are
normalized such that
\begin{equation}
   \widetilde\varphi(0) = \int\limits_0^\infty\!\text{d}\omega\,
   \varphi_\pm(\omega) = 1 \,.
\label{wf6}
\end{equation}
We can formally introduce operators $O_\pm(\omega)$ such that
\begin{equation}
   \langle\,0\,|O_\pm(\omega)|M\rangle
   = f\,\varphi_\pm(\omega)\,\gamma_\pm u \,.
\label{phidef}
\end{equation}
This implies
\begin{eqnarray}
   O_\pm(\omega) &=& \frac{1}{2\pi} \int\text{d}t\,
    \widetilde O_\pm(t)\,e^{i\omega t} 
    = Q^*(0)\,\gamma_\pm\,\delta(iD_+ - \omega)\,q(0) \,,
    \nonumber\\
   \widetilde O_\pm(t) &=& \int\text{d}\omega\,O_\pm(\omega)\,
    e^{-i\omega t} \,.
\label{wf5}
\end{eqnarray}
Expanding in powers of $t$ in the definitions~(\ref{wf4}) and
(\ref{wf5}), we obtain
\begin{equation}
   \widetilde O_\pm(t) = \sum_{n=0}^\infty\,O_\pm^{(n)}\,
   \frac{(-it)^n}{n!} \,,\quad
   \widetilde\varphi_\pm(t) = \sum_{n=0}^\infty\,
   \langle\omega^n\rangle_\pm \frac{(-it)^n}{n!} \,,
\label{Ophiexp}
\end{equation}
where
\begin{eqnarray}
   O_\pm^{(n)} &=& \int\text{d}\omega\,O_\pm(\omega)\,\omega^n
    = Q^*\gamma_\pm (iD_+)^n q \,, \nonumber\\
   \langle\omega^n\rangle_\pm &=& \int\text{d}\omega\,
    \varphi_\pm(\omega)\,\omega^n \,.
\label{wf7}
\end{eqnarray}
Equation~(\ref{phidef}) then implies a relation between the moments
of the momentum-space wave functions and the local,
higher-dimensional operators $O_\pm^{(n)}$:
\begin{equation}
   \langle\,0\,|O_\pm^{(n)}|M\rangle
   = f\,\langle\omega^n\rangle_\pm \gamma_\pm u \,.
\label{moments}
\end{equation}

Using the equations of motion, the first moments of the wave
functions can be calculated in terms of the parameter $\bar\Lambda$
encountered in (\ref{i6})~\cite{subl}. In general, we may write
\begin{equation}
   \langle\,0\,|Q^* iD^\mu q|M\rangle
   = f\,(a v^\mu + b\gamma^\mu) u \,.
\end{equation}
The equations of motion for the light quark, $i\rlap{\,/}{D} q=0$, 
imply that $(a+4 b)=0$. The equations of motion for the heavy quark,
$iv\cdot D Q=0$, can be used to write
\begin{equation}
   \langle\,0\,|Q^* iv\cdot D\,q|M\rangle
   = iv\cdot\partial\,\langle\,0\,|Q^* q|M\rangle
   = \bar\Lambda\,\langle\,0\,|Q^* q|M\rangle \,,
\end{equation}
where $\bar\Lambda=m_M-m_Q$ is the effective mass of the meson $M$ in
the HQET \cite{FNL}. This relation implies that $(a+b)=\bar\Lambda$,
and therefore
\begin{equation}
   \langle\,0\,|Q^* iD^\mu q|M\rangle
   = \case 13 f\bar\Lambda\,(4 v^\mu - \gamma^\mu) u \,.
\label{Dmumat}
\end{equation}
Using this result, we find that the first moments of the wave
functions are given by
\begin{equation}
   \langle\omega\rangle_+ = \case 43 \bar\Lambda \,,\quad
   \langle\omega\rangle_- = \case 23 \bar\Lambda \,.
\label{wf9}
\end{equation}

A similar analysis can be performed for the second moments. Consider
the matrix element
\begin{equation}
   \langle\,0\,|Q^* iD^\mu iD^\nu q|M\rangle
   = f\,\Theta^{\mu\nu} u \,,
\label{theta}
\end{equation}
where the most general form of $\Theta^{\mu\nu}$ is
\begin{equation}
   \Theta^{\mu\nu} = c_1 v^\mu v^\nu + c_2 g^{\mu\nu}
   + c_3 (\gamma^\mu v^\nu + \gamma^\nu v^\mu) 
   + c_4 (\gamma^\mu v^\nu - \gamma^\nu v^\mu)
   + i c_5 \sigma^{\mu\nu} \,.
\label{Thmunu}
\end{equation}
The equations of motion impose three independent relations among the
five parameters $c_i$, which imply that the matrix element in
(\ref{theta}) is completely determined by its antisymmetric
part~\cite{FN}, i.e.\ by the matrix element of the gluon field
$iG^{\mu\nu}=[iD^\mu,iD^\nu]$. For reasons to become clear below, we
find it convenient to introduce two hadronic parameters $\lambda_E^2$
and $\lambda_H^2$ by
\begin{equation}
   c_4 = \case 16 (\lambda_H^2 - \lambda_E^2) \,,\quad
   c_5 = \case 16 \lambda_H^2 \,.
\label{lamdef}
\end{equation}
In terms of these quantities, we obtain
\begin{eqnarray}
   \langle\,0\,|Q^*i G^{\mu\nu} q|M\rangle 
   &=& \case 13 f\,\Big[ (\lambda_H^2 - \lambda_E^2)
    (\gamma^\mu v^\nu - \gamma^\nu v^\mu) + i\lambda_H^2
    \sigma^{\mu\nu} \Big]\,u \,, \nonumber\\ 
   \langle\,0\,|Q^* \case 12 \{ iD^\mu,iD^\nu \}\,q|M\rangle
   &=& \case 13 f\,\Big[ (6\bar\Lambda^2 + 2\lambda_E^2
    + \lambda_H^2) v^\mu v^\nu - (\bar\Lambda^2 + \lambda_E^2
    + \lambda_H^2) g^{\mu\nu} \nonumber\\
   &&\mbox{}- (\bar\Lambda^2 + \case 12 \lambda_E^2)
    (\gamma^\mu v^\nu + \gamma^\nu v^\mu) \Big]\,u \,.
\label{Dsq}
\end{eqnarray}
From the second relation, it follows that the second moments of the
wave functions are given by
\begin{equation}
   \langle\omega^2\rangle_+ = 2\bar\Lambda^2
   + \case 23 \lambda_E^2 + \case 13 \lambda_H^2 \,,\quad
   \langle\omega^2\rangle_- = \case 23 \bar\Lambda^2
   + \case 13 \lambda_H^2 \,.
\label{wf12}
\end{equation}
According to the first equation in (\ref{Dsq}), the moments
$\langle\omega^2\rangle_\pm$ are thus related to normalization
integrals of quark--antiquark--gluon wave functions.

Our definition in (\ref{lamdef}) is such that, in the rest frame of
the heavy meson, the quantities $\lambda_E^2$ and $\lambda_H^2$
parametrize the matrix elements of the chromo-electric and
chromo-magnetic fields, respectively. Defining\footnote{If we define
$D^\mu=(D^0,-\vec{D})$, then $\vec{E}=i[D^0,\vec{D}]$ and
$\vec{H}=-i\vec{D}\times\vec{D}$.}
$E_i=G_{0i}$, $H_i=-\case 12 \epsilon_{ijk} G_{jk}$, and
$\alpha_i=\gamma^0\gamma^i$, we find
\begin{equation}
   \langle\,0\,|Q^* i\vec{\alpha}\cdot\vec{E}\,q|M\rangle
   = f \lambda_E^2 u \,, \quad
   - \langle\,0\,|Q^* \vec{\sigma}\cdot\vec{H}\,q|M\rangle
   = f \lambda_H^2 u \,.
\label{wf10}
\end{equation}

To finish this section, let us switch the heavy-quark spin on and
relate the numerous quark--antiquark wave functions of the
ground-state pseudoscalar and vector mesons, $M$ and $M^*$, to the
HQET wave functions $\varphi_\pm$. These relations are very
conveniently obtained using the covariant tensor formalism described
in Appendix~A. For a pseudoscalar meson $M$, the matrix elements of
the pseudoscalar, axial, and tensor currents are non-zero, and we
define a set of four wave functions in the following
way:\footnote{Contrary to the notation used in the rest of this
paper, here and in (\protect\ref{phidef2}) we use the standard
relativistic normalization of states, which adds a factor
$\sqrt{m_M}$ on the right-hand side of the equations.}
\begin{eqnarray}
   \langle\,0\,|\bar Q(0)\,\gamma_5\,q(z)|M\rangle
   &=& - i f_M m_M\,\widetilde\varphi_P \,, \nonumber\\
   \langle\,0\,|\bar Q(0)\,\gamma^\mu\gamma_5\,q(z)
   |M\rangle &=& f_M [ i\widetilde\varphi_{A1}\,p^\mu
   - m_M\,\widetilde\varphi_{A2}\,z^\mu ] \,, \nonumber\\
   \langle\,0\,|\bar Q(0)\,\sigma^{\mu\nu}\gamma_5\,q(z)|M\rangle
   &=& i f_M\,\widetilde\varphi_T\,(p^\mu z^\nu - p^\nu z^\mu) \,,
\label{phidef1}
\end{eqnarray}
where $\widetilde\varphi_i=\widetilde\varphi_i(p\cdot z) =
\widetilde\varphi_i(m_M t)$. For simplicity, we have omitted the
string operator $E(0,z)$, i.e.\ we have adopted the light-cone gauge
$A_+=0$. In the heavy-quark limit, we obtain, using the results of
Appendix~A:
\begin{equation}
   \widetilde\varphi_P = \frac{\widetilde\varphi_+(t)
   + \widetilde\varphi_-(t)}{2} \,,\quad 
   \widetilde\varphi_{A1} = \widetilde\varphi_+(t) \,,\quad
   \widetilde\varphi_{A2} = \widetilde\varphi_T = \frac{i}{2}\,
   \frac{\widetilde\varphi_+(t) - \widetilde\varphi_-(t)}{t} \,.
\end{equation}
For $t=0$, we obtain in this limit the normalization conditions:
\begin{equation}
   \widetilde\varphi_P(0) = \widetilde\varphi_{A1}(0) = 1 \,,\quad
   \widetilde\varphi_{A2}(0) = \widetilde\varphi_T(0) = 
   \frac{\bar\Lambda}{3} \,,
\end{equation}
where the second relation is a consequence of (\ref{Ophiexp}) and
(\ref{wf9}).

For a vector meson $M^*$ with polarization vector $e$, the matrix
elements of the scalar, vector, axial, and tensor currents are
non-zero, and we introduce a set of six wave functions as follows:
\begin{eqnarray}
   \langle\,0\,|\bar Q(0)\,q(z)|M^*\rangle
   &=& i f_{M^*} m_{M^*}\,\widetilde\varphi_S\,z\cdot e \,,
    \nonumber\\
   \langle\,0\,|\bar Q(0)\,\gamma^\mu\,q(z)|M^*\rangle
   &=& f_{M^*} \Big[ m_{M^*}\,\widetilde\varphi_{V1}\,e^\mu
    +i \widetilde\varphi_{V2}\,(p\cdot z\,e^\mu 
    - e\cdot z\,p^\mu) \Big] \,, \nonumber\\
   \langle\,0\,|\bar Q(0)\,\gamma^\mu\gamma_5\,q(z)|M^*\rangle
   &=& f_{M^*} \widetilde\varphi_A\,
    \epsilon^{\mu\alpha\beta\gamma} z_\alpha p_\beta e_\gamma
    \,, \nonumber\\
   \langle\,0\,|\bar Q(0)\,\sigma^{\mu\nu}\,q(z)|M^*\rangle
   &=& f_{M^*} \Big[ i\widetilde\varphi_{T1}\,
    (e^\mu p^\nu - e^\nu p^\mu) 
    -m_{M^*}\,\widetilde\varphi_{T2}\,(e^\mu z^\nu
    - e^\nu z^\mu) \Big] \,. \nonumber\\
\label{phidef2}
\end{eqnarray}
In the heavy-quark limit, we find the relations
\begin{equation}
   \widetilde\varphi_{V1} = \widetilde\varphi_{T1}
   = \widetilde\varphi_+(t) \,,\quad 
   \widetilde\varphi_S = \widetilde\varphi_{V2}
   = \widetilde\varphi_A = \widetilde\varphi_{T2}
   = \frac{i}{2}\,
   \frac{\widetilde\varphi_+(t) - \widetilde\varphi_-(t)}{t} \,,
\end{equation}
and the corresponding normalization conditions
\begin{equation}
   \widetilde\varphi_{V1}(0) = \widetilde\varphi_{T1}(0) = 1 \,,
   \quad
   \widetilde\varphi_S(0) = \widetilde\varphi_{V2}(0)
   = \widetilde\varphi_A(0) = \widetilde\varphi_{T2}(0)
   = \frac{\bar\Lambda}{3} \,.
\label{norm2}
\end{equation}

The QCD wave functions in momentum space are defined as
\begin{equation}
   \varphi_i(x) = \frac{1}{2\pi} \int\text{d}(p\cdot z)\,
   \widetilde\varphi_i(p\cdot z)\,e^{-ix p\cdot z} \,,
\label{wf16}
\end{equation}
so that $x=\omega/m_M$. On the basis of the behaviour of the
eigenfunctions of the evolution equations as well as sum-rule
inspired arguments, it is usually assumed that for pseudoscalar
mesons $\varphi_{A1}(x)\sim x$ and $\varphi_P(x)\sim 1$ as $x\to
0$~\cite{Chernyak}. For the HQET wave functions, this implies the
behaviour
\begin{equation}
   \varphi_+(\omega) \sim \omega \,,\quad \varphi_-(\omega) \sim 1 
\label{phiasy}
\end{equation} 
as $\omega\to 0$. In Sec.~\ref{SR}, we will indeed find these scaling
laws from an explicit calculation of the wave functions using QCD sum
rules.

\section{Evolution Equations}
\label{Evol}

The definitions of the previous section are somewhat formal, because
the operators involved require renormalization. In this section we
discuss how the ultraviolet divergences in operator matrix elements
can be removed in a consistent way. After doing this, however, we
shall ignore renormalization effects in the further course of this
paper. The reader not interested in the conceptual problem of
renormalization can thus proceed directly with Sec.~\ref{SR}.

We use the $\overline{\text{\sc ms}}$ subtraction scheme in
$d=4-2\varepsilon$ dimensional space-time. The bare and renormalized
operators are related by
\begin{equation}
   O^{\text{bare}}_\pm(\omega) = \int\text{d}\omega'\,
   Z_\pm(\omega,\omega')\,O_\pm(\omega') \,,
\label{ev1}
\end{equation}
where 
\begin{equation}
   Z_\pm(\omega,\omega') = \delta(\omega-\omega')
   - \frac{1}{2\varepsilon}\,z_\pm(\omega,\omega') + \dots \,,
\label{ev2}
\end{equation}
and the ellipses represent poles of higher order in $1/\varepsilon$.
The operators $O_\pm(\omega)$ and hence their matrix elements
$f\varphi_\pm(\omega)$ obey the renormalization-group equations
\begin{equation}
   \frac{\text{d}\,f\varphi_\pm(\omega)}{\text{d}\log\mu}
   + \int\text{d}\omega'\,\Gamma_\pm(\omega,\omega')\,
   f\varphi_\pm(\omega') = 0 \,,
\label{ev3}
\end{equation}
where the anomalous dimensions $\Gamma_\pm(\omega,\omega')$ are
given by
\begin{equation}
   \Gamma_\pm(\omega,\omega') = \alpha_s\,
   \frac{\partial}{\partial\alpha_s}\,z_\pm(\omega,\omega') \,.
\end{equation}
Equation~(\ref{ev3}) is analogous to the Brodsky--Lepage evolution
equation in QCD~\cite{Brodsky}. 

\begin{figure}
\epsfxsize=8cm
\centerline{\epsffile{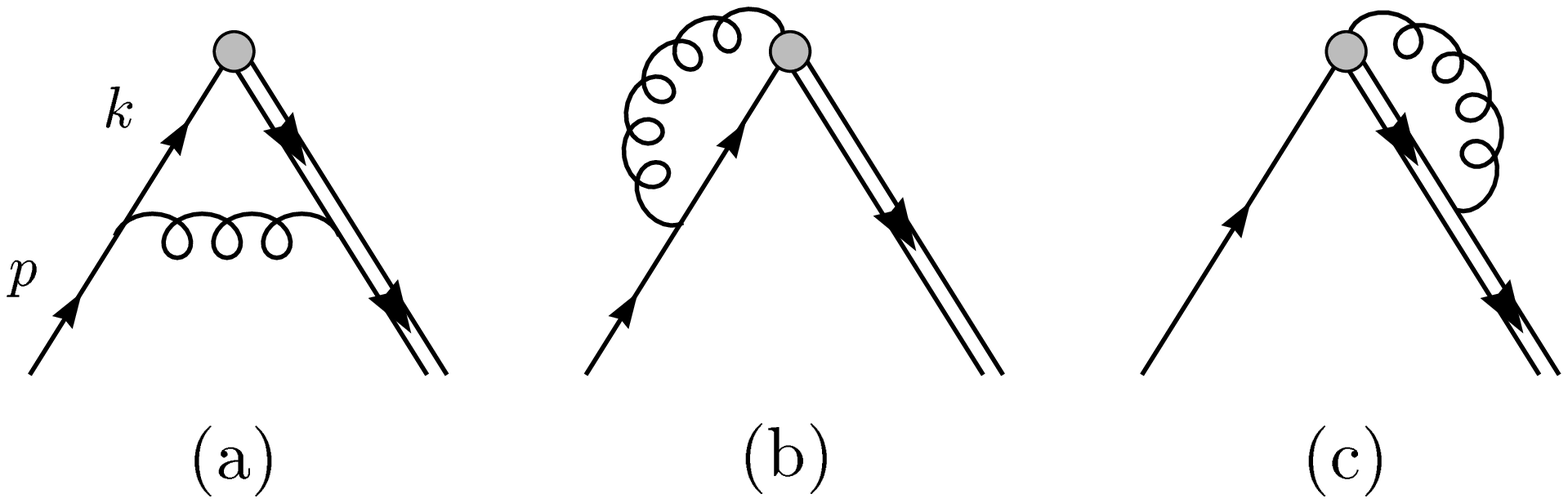}}
\vspace{0.2cm}
\caption{One-loop diagrams contributing to the matrix elements
$\langle\,0\,|O^{\text{bare}}_\pm(\omega)|\omega'\rangle$. The bare
current operators are represented by a circle.}
\label{Fevol}
\end{figure}

\begin{figure}
\epsfxsize=8cm
\centerline{\epsffile{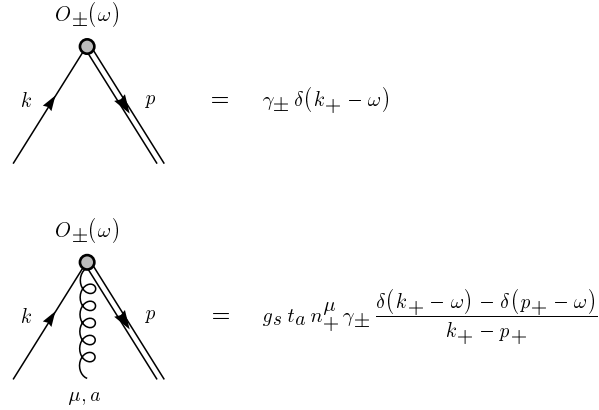}}
\vspace{0.2cm}
\caption{Feynman rules for vertices involving $O_\pm(\omega)$.}
\label{rules}
\end{figure}

In order to obtain the anomalous dimensions
$\Gamma_\pm(\omega,\omega')$ at one-loop order, we consider the
matrix elements $\langle\,0\,|O^{\text{bare}}_\pm(\omega)|
\omega'\rangle$, where $|\omega'\rangle$ represents a state
consisting of a scalar heavy antiquark at rest and a light quark with
momentum $p_+=\omega'$. According to~(\ref{ev1}), these matrix
elements equal $Z_\pm(\omega,\omega')\,\gamma_\pm u$. The relevant
one-particle irreducible loop diagrams are shown in Fig.~\ref{Fevol}.
Although the operators $O_\pm(\omega)$ in (\ref{wf5}) take a
particularly simple form in the light-cone gauge $A_+=0$, we refrain
from adopting such a singular gauge and work instead in the Feynman
gauge.\footnote{We have repeated the calculation of
$\Gamma_+(\omega,\omega')$ in the light-cone gauge and obtained the
same result as in the Feynman gauge; however, we could not find an
easy way to recover the correct result for
$\Gamma_-(\omega,\omega')$.}
To obtain the Feynman rules for vertices involving the operators
$O_\pm(\omega)$, we start from their definition as the Fourier
transforms of the non-local operators $\widetilde O_\pm(t)$ and
obtain, to first order in the gauge field:
\begin{eqnarray}
   O_\pm(\omega) &=& \frac{1}{2\pi}\int\text{d}t\,e^{i\omega t}
    \bigg\{ Q^*(0)\,\gamma_\pm\,e^{t\partial_+}\,q(0) \nonumber\\
   &&\mbox{}- i\int\text{d}\tau\,Q^*(0)\,\gamma_\pm\,
    e^{\tau\partial_+}\,A_+(0)\,e^{(t-\tau)\partial_+}\,q(0)
    + \dots \bigg\} \,.
\end{eqnarray}
It is then straightforward to derive the Feynman rules shown in Fig.~\ref{rules}. 

Let us then sketch the calculation of the diagrams in
Fig.~\ref{Fevol}. The contribution of the first diagram is:
\widetext
\begin{equation}
   -i C_F\,\frac{\alpha_s}{2\pi}\,\mu^{2\varepsilon}
   \int \text{d}k_+ \text{d}k_-\,
   \frac{\text{d}^{2-2\varepsilon}k_\bot}{(2\pi)^{2-2\varepsilon}}\,
   \frac{\delta(k_+-\omega)\,\gamma_\pm\,\rlap/k\,u}
   {(k^0-i0)(k^2+i0)[(k-p)^2+i0]} \,,
\end{equation}
where $C_F=(N_c^2-1)/(2 N_c)$. The $k_+$ integral is trivial to
perform, and the $k_-$ integral is readily calculated by the method
of residues. The poles of the integrand are located at
$k_-=-\omega+i0$, $k_-=(\vec k_\bot^2-i0)/\omega$, and $k_-=(\vec
k_\bot^2-i0)/ (\omega-\omega')$. If $\omega<0$, all poles lie in the
upper half plane, and the integral vanishes. For $\omega>0$, it is
necessary to distinguish the cases $\omega>\omega'$ and
$\omega<\omega'$. For $\omega>\omega'$, we close the contour in the
upper half plane and set $k_-=-\omega$. For $\omega<\omega'$, we
close the contour in the lower half plane and set $k_-=\vec
k_\bot^2/\omega$. Only in the second case and for the minus
projection there is an ultraviolet divergence, which arises from the
$k_\bot$ integration. Keeping only the singular term, we obtain for
the contribution to $Z_\pm(\omega,\omega')$:
\begin{equation}
   C_F\,\frac{\alpha_s}{4\pi\varepsilon}\,(1\mp 1)\,
   \frac{1}{\omega'}\,\theta(\omega'-\omega) \,.
\end{equation}
The other two diagrams in Fig.~\ref{Fevol} are evaluated in a similar
way. For the second one, we find an ultraviolet divergence for the
plus projection if $k_+<\omega'$. Its contribution to
$Z_\pm(\omega,\omega')$ is:
\begin{eqnarray}
   && - C_F\,\frac{\alpha_s}{4\pi\varepsilon}\,(1\pm 1)
    \int\text{d}k_+\,\frac{k_+}{\omega'}\,
    \frac{\delta(k_+-\omega) - \delta(\omega'-\omega)}{k_+-\omega'}
    \nonumber\\
   &=& - C_F\,\frac{\alpha_s}{4\pi\varepsilon}\,(1\pm 1)\,\left\{
    \left[ \frac{1}{\omega'} + \frac{1}{(\omega-\omega')_+} \right]\,
    \theta(\omega'-\omega) - \delta(\omega'-\omega) \right\} \,.
\end{eqnarray}
The distribution $1/(\omega-\omega')_+$ is defined such that
\begin{equation}
   \int\text{d}\omega\,\frac{f(\omega)}{(\omega-\omega')_+}
   = \int\text{d}\omega\,\frac{f(\omega)-f(\omega')}
   {\omega-\omega'}
\label{ev10}
\end{equation}
for any smooth function $f(\omega)$. The third diagram has an
ultraviolet divergence if $k_+>\omega'$. Its contribution is:
\begin{equation}
   C_F\,\frac{\alpha_s}{2\pi\varepsilon}\,
   \frac{1}{(\omega-\omega')_+}\,\theta(\omega-\omega') \,.
\end{equation}
Finally, we have to add the contributions from the wave-function
renormalization of the external lines, which gives~\cite{PoWi}
\begin{equation}
   Z_Q^{1/2} Z_q^{1/2}\,\delta(\omega-\omega')
   = \left( 1 + C_F\,\frac{\alpha_s}{8\pi\varepsilon} \right)
   \delta(\omega-\omega') \,.
\end{equation}
Collecting all terms, we obtain for the quantities
$z_\pm(\omega,\omega')$ defined in (\ref{ev2}):
\begin{eqnarray}
   z_\pm(\omega,\omega') &=& C_F\,\frac{\alpha_s}{\pi}\,
    \Bigg\{ \pm \frac{\theta(\omega'-\omega)}{\omega'}
    - \frac{\theta(\omega-\omega')}{(\omega-\omega')_+}
    - \frac{3}{4}\,\delta(\omega-\omega') \nonumber\\
   &&\mbox{}+ \frac 12 (1\pm 1)
    \left[ \frac{\theta(\omega'-\omega)}{(\omega-\omega')_+}
    - \delta(\omega-\omega') \right] \Bigg\} \,.
\label{ev8}
\end{eqnarray}
To one-loop order, the anomalous dimensions are given by the same
expression. We thus obtain
\begin{eqnarray}
   \Gamma_+(\omega,\omega') &=& C_F\,\frac{\alpha_s}{\pi}
    \left[ -\frac{1}{|\omega-\omega'|_+}
    + \frac{\theta(\omega'-\omega)}{\omega'}
    -\frac 54 \delta(\omega-\omega') \right] \,, \nonumber\\
   \Gamma_-(\omega,\omega') &=& C_F\,\frac{\alpha_s}{\pi}
    \left[ -\frac{\theta(\omega-\omega')}{(\omega-\omega')_+}
    - \frac{\theta(\omega'-\omega)}{\omega'}
    -\frac 14 \delta(\omega-\omega') \right] \,.
\label{ev9b}
\end{eqnarray}
\narrowtext\noindent

In the familiar case of QCD wave functions for light-quark systems,
the renormalization of the non-local operators analogous to
$O_\pm(\omega)$ would suffice to renormalize the tower of local
operators defined analogous to $O_\pm^{(n)}$ in (\ref{wf7}).
Accordingly, the renormalization of the wave functions $\varphi_i(x)$
renders their moments $\langle x^n\rangle_i$ finite. Unfortunately,
this situation does not hold in the case of wave functions for heavy
mesons defined in the HQET. The reason for this unexpected fact is
that in the HQET the wave functions depend on the dimensional
variable $\omega$, which takes values between 0 and $\infty$. As a
consequence, equation~(\ref{wf7}), which defines the operators
$O_\pm^{(n)}$ in terms of weighted integrals of $O_\pm(\omega)$, does
not hold for the renormalized operators; the integral over $\omega$
has an additional ultraviolet divergence not yet removed by the
renormalization of $O_\pm(\omega)$. This divergence must be removed
separately.

Consider, as an example, the simplest case $n=0$. Then the bare
operators $O_\pm^{(0),\text{bare}} = Q^*\gamma_\pm q$ are local
heavy-light current operators with dimension three, which are
renormalized in the following way:
\begin{equation}
   O_\pm^{(0),\text{bare}} = Z_0\,O_\pm^{(0)}
   = \left( 1 - \frac{1}{2\varepsilon}\,\Gamma_0 + \dots \right)
   O_\pm^{(0)} \,,
\label{Z0}
\end{equation}
where
\begin{equation}
   \Gamma_0 = -\frac 34\,C_F\,\frac{\alpha_s}{\pi}
\label{hybrid}
\end{equation}
is the well-known one-loop hybrid anomalous
dimension~\cite{OneLoop,Vol1,PoWi}. On the other hand, the bare
operators can be expressed in terms of integrals over the
renormalized operators $O_\pm(\omega)$ using (\ref{ev1}). This gives:
\begin{equation}
   O_\pm^{(0),\text{bare}} = \int\text{d}\omega\,\text{d}\omega'\,
   Z_\pm(\omega,\omega')\,O_\pm(\omega') 
   = \left( \int\text{d}\omega\,Z_\pm(\omega,\omega') \right)
   \int\text{d}\omega\,O_\pm(\omega) \,.
\label{Zint}
\end{equation}
As mentioned above, the last integral over the renormalized operators
$O_\pm(\omega)$ is ultraviolet divergent from the region
$\omega\to\infty$, and we define
\begin{equation}
   \int\text{d}\omega\,O_\pm(\omega) = Z'_\pm\,O_\pm^{(0)} 
   = \left( 1 - \frac{1}{2\varepsilon}\,\Gamma'_\pm + \dots \right)
   O_\pm^{(0)} \,.
\label{ev4}
\end{equation}
Combining (\ref{Z0}), (\ref{Zint}) and (\ref{ev4}), we find that
\begin{equation}
   Z_0 = Z'_\pm \int\text{d}\omega\,Z_\pm(\omega,\omega') \,,\quad
   \Gamma_0 = \Gamma'_\pm + \int\text{d}\omega\,
   \Gamma_\pm(\omega,\omega') \,.
\label{ev5}
\end{equation}
With our explicit expressions in (\ref{ev9b}), we obtain
\begin{equation}
   Z'_\pm = 1\pm C_F\,\frac{\alpha_s}{4\pi\varepsilon} \,,\quad
  \Gamma'_\pm = \mp C_F\,\frac{\alpha_s}{2\pi} \,.
\end{equation}

To see in detail how this works, we calculate the ultraviolet
divergences of the matrix elements
$\langle\,0\,|O_\pm^{(0),\text{bare}}|\omega'\rangle$ to one-loop
order, leaving the integration over $k_+$ until the end. Since now we
are dealing with local operators, the only one-particle irreducible
diagram is a vertex diagram analogous to that in Fig.~\ref{Fevol}a.
Its contribution is:
\widetext
\begin{eqnarray}
   2 C_F\,\alpha_s\gamma_\pm\,u\,\mu^{2\varepsilon}
   &\Bigg\{& \pm \int\limits_{\omega'}^\infty\!\text{d}\omega\,
    \omega \int\frac{\text{d}^{2-2\varepsilon}k_\bot}
   {(2\pi)^{2-2\varepsilon}}\,
   \frac{1}{(\vec k_\bot^2 + \omega^2 -i0)
    \big[\vec k_\bot^2 + \omega(\omega-\omega') -i0 \big]}
    \nonumber\\
   &&\mbox{}+ \frac{1}{\omega'} \int\limits_0^{\omega'}\!
    \text{d}\omega\,\omega\, \int
    \frac{\text{d}^{2-2\varepsilon}k_\bot}{(2\pi)^{2-2\varepsilon}}\,
    \frac{k_\pm}{(\vec k_\bot^2 + \omega^2 -i0)(\vec k_\bot^2 -i0)}
    \Bigg\} \,, 
\end{eqnarray}
where in the numerator of the second integral we have to substitute
$k_-=\vec k_\bot^2/\omega$. This expression shows the origin of the
two types of ultraviolet divergences. The $k_\bot$ integral in the
first term is convergent, but there is a logarithmic divergence in
the remaining integral over $\omega$. In the second term, the
$\omega$ integral is cut off in the ultraviolet; however, the
$k_\bot$ integral diverges for the minus projection. Keeping only the
poles in $1/\varepsilon$, we obtain
\begin{equation}
   C_F\,\frac{\alpha_s}{2\pi}\,\gamma_\pm\,u\,\Bigg\{
   \pm \omega'^{2\varepsilon} \int\limits_{\omega'}^\infty\!
   \text{d}\omega\,\omega^{-1-2\varepsilon} + (1\mp 1)\,
   \frac{1}{2\varepsilon} + \dots \Bigg\}
   = C_F\,\frac{\alpha_s}{4\pi\varepsilon}\,\gamma_\pm\,u \,.
\label{UVs}
\end{equation}
\narrowtext\noindent
Adding to this the contributions from the wave-function
renormalization of the external lines, we recover the expressions for
$Z_0$ and $\Gamma_0$ given in (\ref{Z0}) and (\ref{hybrid}). The
contribution in (\ref{UVs}) arising from the logarithmic divergence
of the $\omega$ integral is removed by the factor $Z'_\pm$ defined in
(\ref{ev4}), whereas all other contributions are removed by the
renormalization of the operators $O_\pm(\omega)$.

\section{QCD Sum Rules}
\label{SR}

The first moments of the wave functions $\varphi_\pm(\omega)$ are
known exactly from the equations of motion, as shown in~(\ref{wf9}).
However, they only set the scale of $\omega$. In order to obtain
information about the shape of the wave functions, we need to
consider some of the higher moments. In this section, we use QCD sum
rules to investigate the parameters $\lambda_E^2$ and $\lambda_H^2$,
which according to (\ref{wf12}) determine the second moments of
$\varphi_\pm(\omega)$.

Let us consider the operators
\begin{eqnarray}
   O_E &=& Q^* i\vec{\alpha}\cdot\vec{E}\,q
    = v_\nu v^\alpha\,Q^* G^{\mu\nu}\sigma_{\mu\alpha}\,q \,,
    \nonumber\\
   O_H &=& - Q^* \vec{\sigma}\cdot\vec{H}\,q 
    = \left( \case 12 g_\nu^\alpha - v_\nu v^\alpha \right)
    Q^* G^{\mu\nu}\sigma_{\mu\alpha}\,q \,,
\label{sr1}
\end{eqnarray}
whose matrix elements give $f\lambda_E^2$ and $f\lambda_H^2$,
respectively. To obtain the sum rule for $\lambda_E^2$ (the sum rule
for $\lambda_H^2$ is obtained in a similar way), we investigate the
correlator
\begin{equation}
   \langle\,0\,| T\{ O_E(x), \bar q\,\case 12(1+\rlap/v) Q(0) \}
   |\,0\,\rangle = \frac{1+\rlap/v}{2}\,\theta(v\cdot x)\,
   \delta(\vec x_\bot)\,\Pi_E(v\cdot x) \,,
\label{sr2}
\end{equation}
where $\vec x_\bot$ contains the components of $x$ orthogonal to $v$.
The operator $\bar q\,\case 12(1+\rlap/v) Q$ has the quantum numbers
of the ground-state meson. Next, we analytically continue
$\Pi_E(v\cdot x)$ to $v\cdot x=-i\tau$.\footnote{The procedure of
analytic continuation of the coordinate-space correlator to imaginary
time is equivalent to the Borel transformation of the corresponding
momentum-space correlator. Thus, our sum rules coincide with the
usual Laplace sum rules.}
The correlator can be expressed via its spectral density
$\rho_E(\omega)$:
\begin{equation}
   \Pi_E(\tau) = \int\limits_0^\infty\!\text{d}\varepsilon\,
   \rho_E(\varepsilon)\,e^{-\varepsilon\tau} \,,
\label{sr3}
\end{equation}
which contains the contribution $f^2\lambda_E^2
\delta(\varepsilon-\bar\Lambda)$ from the ground-state meson, as well
as contributions from excited states, which we will refer to as the
continuum. Hence, the phenomenological expression for the correlator
is
\begin{equation}
   \Pi_E(\tau) = f^2\lambda_E^2 e^{-\bar\Lambda\tau}
   + \Pi_E^{\text{cont}}(\tau) \,.
\label{sr4}
\end{equation}
For sufficiently small values of $\tau$, the correlator can be
calculated in QCD using the operator product expansion~\cite{SVZ}.
The theoretical spectral density, $\rho_E^{\text{th}}(\varepsilon)$,
contains perturbative as well as non-perturbative contributions,
where the latter are proportional to vacuum condensates of local,
gauge-invariant operators. The continuum contributions in (\ref{sr4})
are usually modeled by the theoretical spectral density above a
threshold $\varepsilon_c$, which is called the continuum threshold.
Equating the two expressions for $\Pi_E(\tau)$ obtained in this way,
we derive the sum rule
\begin{equation}
   f^2\lambda_E^2 e^{-\bar\Lambda\tau}
   = \int\limits_0^{\varepsilon_c}\!\text{d}\varepsilon\,
   \rho_E^{\text{th}}(\varepsilon)\,e^{-\varepsilon\tau} \,.
\label{sr5}
\end{equation}
A similar sum rule holds for the product $f^2\lambda_H^2$.
 
\begin{figure}
\epsfxsize=6cm
\centerline{\epsffile{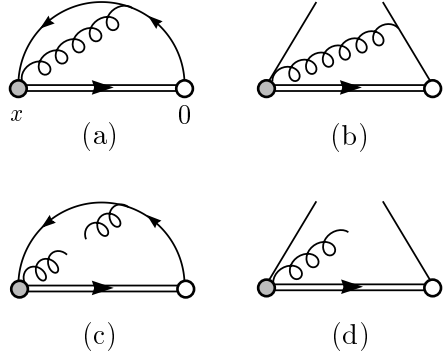}}
\caption{Non-vanishing diagrams for the correlators $\Pi_E$ and
$\Pi_H$. The higher-dimensional current operators $O_E$ and $O_H$ are
represented by a gray circle; the interpolating current is
represented by a white circle.}
\label{Fsr}
\end{figure}

The leading contributions in the operator product expansion of the
correlators are shown in Fig.~\ref{Fsr}. We include the leading
perturbative contribution, as well as the contributions of the quark
condensate $\langle\bar q q\rangle$ (dimension $d=3$), the gluon
condensate $\langle G^2\rangle=\langle G_{a\mu\nu}
G_a^{\mu\nu}\rangle$ ($d=4$), and the mixed quark--gluon condensate
$\langle\bar q\,\sigma_{\mu\nu} G^{\mu\nu} q\rangle\equiv
m_0^2\langle\bar q q\rangle$ ($d=5$). In the calculation of the
non-per\-tur\-ba\-tive contributions, it is convenient to use the
fixed-point gauge $x_\mu A^\mu(x)=0$; then the heavy quark does not
interact with gluons. After a straightforward calculation, we obtain
the sum rules:
\begin{eqnarray}
   f^2\lambda_E^2 e^{-\bar\Lambda\tau} &=& - N_c C_F\,
    \frac{\alpha_s}{\pi^3\tau^5}\,\delta_4(\varepsilon_c\tau)
    - \frac{m_0^2\langle\bar q q\rangle}{16} \,, \nonumber\\
   f^2\lambda_H^2 e^{-\bar\Lambda\tau} &=& - N_c C_F\, 
    \frac{\alpha_s}{2\pi^3\tau^5}\,\delta_4(\varepsilon_c\tau)
    - C_F\,\frac{3\alpha_s}{4\pi\tau^2}\,
    \langle\bar q q\rangle\,\delta_1(\varepsilon_c\tau)
    \nonumber\\
   &&\mbox{}+ \frac{\alpha_s\langle G^2\rangle}{16\pi\tau}\,
    \delta_0(\varepsilon_c\tau) 
    - \frac{m_0^2\langle\bar q q\rangle}{16} \,,
\label{sr9}
\end{eqnarray}
where
\begin{equation}
   \delta_n(x) = \theta(x) \left( 1 - e^{-x} \sum_{m=0}^n
   \frac{x^m}{m!} \right) \,.
\label{sr10}
\end{equation}
The fact that the sum rule for $\lambda_E^2$ does not contain
contributions from the quark and gluon condensates is a consequence
of the fact that, in the fixed-point gauge, the light quark interacts
only with the magnetic components of the gluon field \cite{TechRev}.
In Appendix~B, we also list the contributions of higher-dimensional
quark--gluon condensates ($d=7$) to the sum rules. For the range of
$\tau$ values considered below, we expect that the contributions of
these condensates are very small. Since their values are moreover
unknown, they will be neglected in the numerical analysis.

\begin{figure}
\epsfxsize=7cm
\centerline{\epsffile{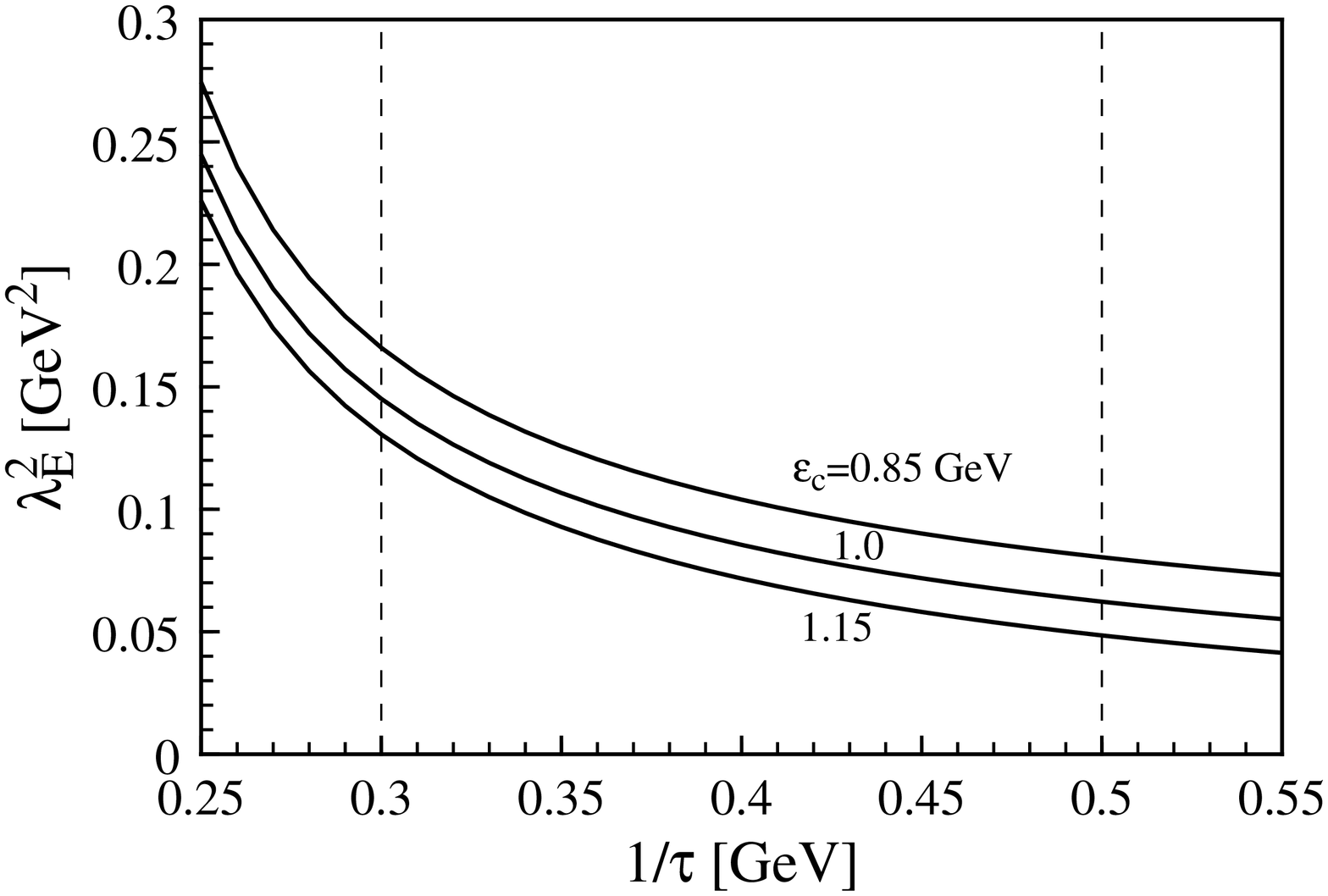}}
\epsfxsize=7cm
\centerline{\epsffile{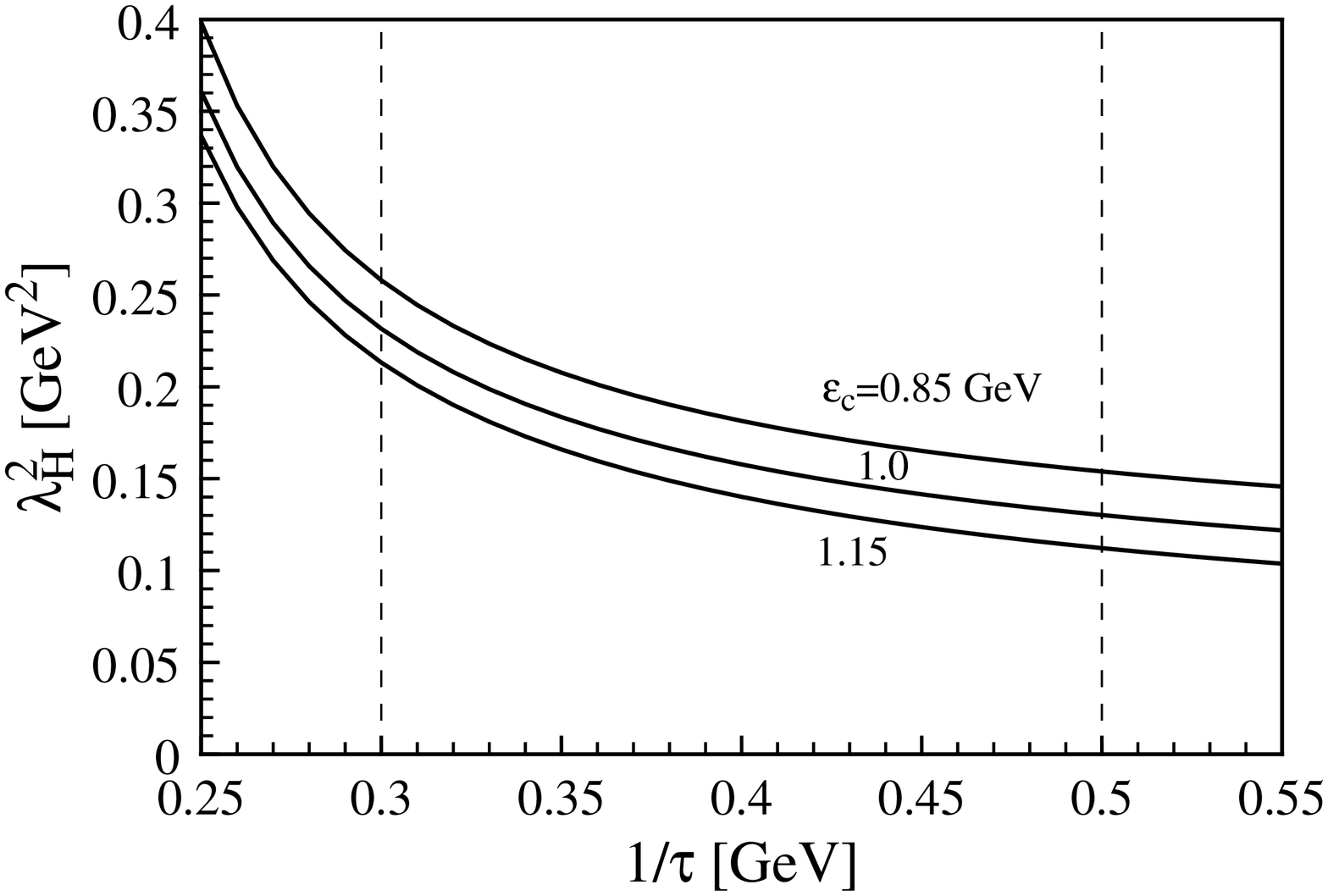}}
\vspace{0.2cm}
\caption{Sum-rule results for $\lambda_E^2$ (upper plot) and
$\lambda_H^2$ (lower plot) as a function of $1/\tau$, for three
values of the continuum threshold $\varepsilon_c$. The stability
window lies in between the dashed lines.}
\label{Fsres}
\end{figure}

It is convenient to divide the sum rules in (\ref{sr9}) by the sum
rule for the meson decay constant derived in~\cite{Shuryak,N1,BBG}:
\begin{equation}
   f^2 e^{-\bar\Lambda\tau} = \frac{N_c}{2\pi^2\tau^3}\,
   \delta_2(\varepsilon_c\tau) - \frac{\langle\bar q q\rangle}{4}
   \left( 1 - \frac{m_0^2\tau^2}{16} \right) \,.
\label{f2sum}
\end{equation}
This procedure leads to expressions for $\lambda_E^2$ and
$\lambda_H^2$ as a function of $\tau$, the continuum threshold
$\varepsilon_c$, and the vacuum condensates. For our analysis we use
the standard values~\cite{SVZ}
\begin{equation}
   \langle\bar q q\rangle = - (0.23~\text{GeV})^3 \,, \quad
   \alpha_s \langle G^2\rangle = 0.04~\text{GeV}^4 \,, \quad
   m_0^2 = 0.8~\text{GeV}^2 \,,
\end{equation}
as well as $\alpha_s=0.4$. The value of the continuum threshold
extracted from the analysis of the sum rule (\ref{f2sum}) is
$\varepsilon_c=1.00\pm 0.15$~GeV \cite{N1,BBG}. Moreover, stability
of this sum rule requires that the parameter $\tau$ be in the range
$0.3~\text{GeV}<1/\tau<0.5~\text{GeV}$, which is called the stability
window. Our numerical results for $\lambda_E^2$ and $\lambda_H^2$ as
a function of $1/\tau$ are shown in Fig.~\ref{Fsres}. We observe a
sizeable dependence of the results on the parameter $\tau$, as it is
not unexpected for the matrix elements of higher-dimensional local
operators such as $O_E$ and $O_H$. Nevertheless, taking an
average over the stability window determined from the sum rule
(\ref{f2sum}), we obtain as a rough estimate:
\begin{equation}
   \lambda_E^2 = (0.11\pm 0.06)~\text{GeV}^2 \,,\quad
   \lambda_H^2 = (0.18\pm 0.07)~\text{GeV}^2 \,.
\end{equation}
In units of $\bar\Lambda\approx 0.55$~GeV \cite{Patricia,subl}, this
implies $\lambda_E^2/\bar\Lambda^2=0.36\pm 0.20$ and
$\lambda_H^2/\bar\Lambda^2=0.60\pm 0.23$.

\begin{figure}
\epsfxsize=8cm
\centerline{\epsffile{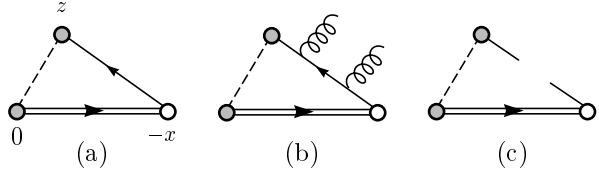}}
\vspace{0.2cm}
\caption{Leading diagrams for the correlators of the bilocal
operators $\widetilde O_\pm(t)$ (gray circles) with the local current
$\bar q Q$ (white circle).}
\label{Fsr2}
\end{figure}

According to (\ref{wf12}), the parameters $\lambda_E^2$ and
$\lambda_H^2$ determined, together with $\bar\Lambda^2$, the second
moments of the meson wave functions. In order to learn more about the
shape of the wave functions, we shall now go a step further and
construct sum rules for the functions $\varphi_\pm(\omega)$
themselves. To this end, we start from the correlator\footnote{
Expanding this correlator in powers of $t$, one recovers, according
to (\protect\ref{Ophiexp}) and (\protect\ref{moments}), the sum rules
for the moments $f^2\langle\omega^n\rangle_\pm$. To lowest order,
this gives the sum rule for the decay constant $f$ considered in
\protect\cite{Shuryak,N1,BBG}. The sum rule for
$f^2\langle\omega\rangle_+$ has been considered
in~\protect\cite{CZZ}; however, the mixed-condensate contribution had
the wrong sign. Note that, because of (\protect\ref{wf9}), the sum
rules for $f^2\langle\omega\rangle_\pm$ can be obtained by taking
derivatives with respect to $\tau$ in the sum rule for $f^2$. Taking
linear combinations of the sum rules for
$f^2\langle\omega^2\rangle_\pm$, we recover the sum rules for
$\lambda_E^2$ and $\lambda_H^2$.}
\begin{equation}
   \langle\,0\,| T\{ \widetilde O_\pm(t), \bar q\,
   \case 12(1+\rlap/v) Q(-x) \} |\,0\,\rangle 
   = \gamma_\pm\,\frac{1+\rlap/v}{2}\,\theta(v\cdot x)\, 
   \delta(\vec x_\bot)\,\widetilde\Pi_\pm(v\cdot x,t) \,,
\label{sr12}
\end{equation}
where $\widetilde O_\pm(t)$ has been defined in~(\ref{Opmt}), and
analytically continue $\widetilde\Pi_\pm(v\cdot x,t)$ to $v\cdot
x=-i\tau$. Using the light-quark propagator in coordinate space
\cite{TechRev}, it is straightforward to obtain the contribution of
the diagrams in Fig.~\ref{Fsr2}a,b. The result is:
\begin{eqnarray}
   \widetilde\Pi_+^{(1)}(\tau,t)
   &=& \frac{N_c}{2\pi^2\tau(\tau+2it)^2}
    \left( 1 - \frac{\pi\alpha_s\langle G^2\rangle\tau^2 t^2}
    {48 N_c} \right) \,, \nonumber\\
   \widetilde\Pi_-^{(1)}(\tau,t)
   &=& \frac{N_c}{2\pi^2\tau^2(\tau+2it)}
    \left( 1 - \frac{\pi\alpha_s\langle G^2\rangle\tau^2 t^2}
    {48 N_c} \right) \,.
\label{sr13}
\end{eqnarray}
The contribution with the cut light-quark line shown in
Fig.~\ref{Fsr2}c involves the trilocal, non-collinear quark
condensate~\cite{Higher}. This object is discussed in detail in
Appendix~B. Keeping for simplicity only the leading term in the
operator product expansion, we obtain
\begin{equation}
   \widetilde\Pi_\pm^{(2)}(\tau,t) = -\case 14\,
   \langle\bar q q\rangle \int\text{d}\nu\,\widetilde f_S(\nu)\,
   e^{-\nu\tau(\tau+2it)} \,,
\label{Piqq}
\end{equation}
where $\widetilde f_S(\nu)$ describes the distribution of quarks with
virtuality $\nu$ in the vacuum. Unfortunately, little is known about
the shape of this function. A simple ansatz is discussed in
Appendix~B.

Next we perform the Fourier transform of the correlator
in~(\ref{sr12}) with respect to $t$, which leads to
\begin{equation}
   \langle\,0\,| T\{ O_\pm(\omega), \bar q\,
   \case 12(1+\rlap/v) Q(-x) \}|\,0\,\rangle 
   = \gamma_\pm\,\frac{1+\rlap/v}{2}\,\theta(v\cdot x)\,
   \delta(\vec x_\bot)\,\Pi_\pm(v\cdot x,\omega) \,.
\label{sr18}
\end{equation}
The leading perturbative terms in~(\ref{sr13}) give
\begin{equation}
   \Pi_+^{(1)}(\tau,\omega) = \frac{N_c}{8\pi^2\tau}\,\omega\,
   e^{-\omega\tau/2} \,,\quad
   \Pi_-^{(1)}(\tau,\omega) = \frac{N_c}{4\pi^2\tau^2}\,
   e^{-\omega\tau/2} \,.
\label{sr19}
\end{equation}
The contributions proportional to the gluon condensate lead to
singular behaviour of the form $\delta(\omega)$ and
$\delta'(\omega)$. These terms would acquire a finite width if the
non-locality of the gluon condensate would be taken into
account~\cite{Mikh2}. Below, we shall neglect the gluon condensate.
The Fourier transform of the quark-condensate contribution
in~(\ref{Piqq}) is given by
\begin{equation}
   \Pi_\pm^{(2)}(\tau,\omega)
   = -\frac{\langle\bar q q\rangle}{8\tau}\,
   \widetilde f_S\left(\frac{\omega}{2\tau}\right)\,
   e^{-\omega\tau/2} \,,
\label{sr24}
\end{equation}
i.e.\ it is directly determined by the virtuality distribution of
quarks in the vacuum.

\begin{figure}
\epsfxsize=7cm
\centerline{\epsffile{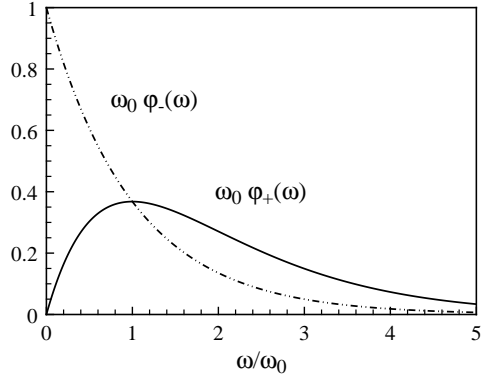}}
\caption{Model wave functions $\varphi_\pm(\omega)$ defined in
(\protect\ref{sr26}).}
\label{Fmod}
\end{figure}

In order to perform the continuum subtraction for the perturbative
contributions, we calculate the inverse Laplace transforms of the
expressions in (\ref{sr19}), which give the corresponding spectral
densities. We finally arrive at the sum rules
\begin{eqnarray}
   f^2\varphi_+(\omega)\,e^{-\bar\Lambda\tau}
   &=& \frac{N_c}{8\pi^2\tau}\,\omega\,e^{-\omega\tau/2}\,
    \delta_0\left[ \left(\varepsilon_c-\case{\omega}{2}\right)\tau
    \right] - \frac{\langle\bar q q\rangle}{8\tau}\,
    \widetilde f_S\left(\frac{\omega}{2\tau}\right)\,
    e^{-\omega\tau/2} \,, \nonumber\\
   f^2\varphi_-(\omega)\,e^{-\bar\Lambda\tau}
   &=& \frac{N_c}{4\pi^2\tau^2}\,e^{-\omega\tau/2}\,
    \delta_1\left[ \left(\varepsilon_c-\case{\omega}{2}\right)\tau
    \right] - \frac{\langle\bar q q\rangle}{8\tau}\,
    \widetilde f_S\left(\frac{\omega}{2\tau}\right)\,
    e^{-\omega\tau/2} \,.
\label{sr25}
\end{eqnarray}
The contributions from the non-local quark condensate fall off
quickly for both $\omega\to 0$ and $\omega\to\infty$. The first
statement follows from a general property of the function $\widetilde
f_S$ (see Appendix~B). However, the precise functional form of these
contributions is unknown. The leading perturbative contributions
vanish for $\omega>2\varepsilon_c$. They suggest the model shapes
\begin{equation}
   \varphi_+(\omega) = \frac{\omega}{\omega_0^2}\,
   e^{-\omega/\omega_0} \,,\quad
   \varphi_-(\omega) = \frac{1}{\omega_0}\,e^{-\omega/\omega_0} \,,
\label{sr26}
\end{equation}
which do indeed exhibit the correct behaviour for $\omega\to 0$
[cf.~(\ref{phiasy})]. These functions are shown in Fig.~\ref{Fmod}.
The parameter $\omega_0=\frac 23\bar\Lambda$ is fixed by~(\ref{wf9}).
In this simple model, we obtain for the second moments
$\langle\omega^2\rangle_+ = 3\langle\omega^2\rangle_- = \case 83
\bar\Lambda^2$. This corresponds to $\lambda_E^2=\lambda_H^2=\frac
23\bar\Lambda^2$, which does not contradict our sum-rule estimates
obtained earlier in this section.

\section{Asymptotics of Form Factors}
\label{IW2}

\begin{figure}
\epsfxsize=7cm
\centerline{\epsffile{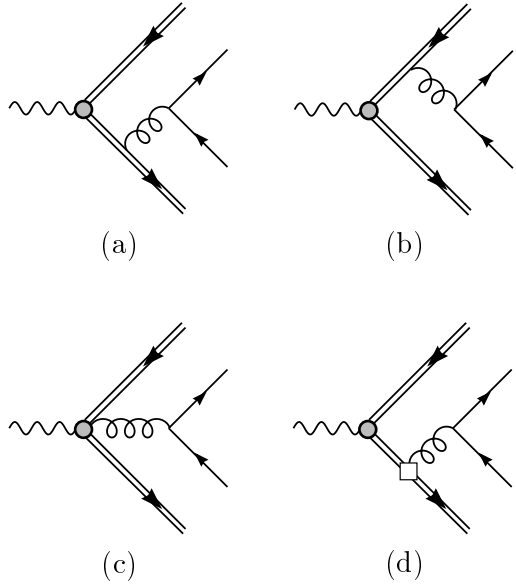}}
\vspace{0.2cm}
\caption{Hard-gluon exchange contributions to heavy-meson form
factors. The external current is represented by the wave line
attached to the gray circle. The white square in (d) represents an
insertion of $1/m_Q$-suppressed operators from the effective
Lagrangian of the HQET.}
\label{FIW}
\end{figure}

We shall now use the results obtained in the previous sections to
analyse, in a model-independent way, the asymptotic behaviour at
large recoil of the form factors describing the current matrix
elements between two heavy mesons. The contribution of the
quark--antiquark wave functions to the Isgur--Wise form factor is
depicted by the diagrams in Fig.~\ref{FIW}a,b. To deal with two heavy
mesons moving at different velocities $v$ and $v'$, it is convenient
to choose the Breit frame, in which the two mesons move in opposite
directions with rapidities $\pm\vartheta/2$, so that
\begin{equation}
   v^\mu = (\cosh\frac{\vartheta}{2},0,0,
    \sinh\frac{\vartheta}{2}) \,, \quad
   v^{\prime\mu} = (\cosh\frac{\vartheta}{2},0,0,
   -\sinh\frac{\vartheta}{2}) \,, 
\end{equation}
and $v\cdot v' = \cosh\vartheta$. In terms of the light-cone vectors
$n_\pm^\mu$, we have
\begin{equation}
   v^\mu + v^{\prime\mu} = \cosh\frac{\vartheta}{2}\,
  (n_+^\mu + n_-^\mu) \,, \quad
   v^\mu - v^{\prime\mu} = -\sinh\frac{\vartheta}{2}\,
  (n_+^\mu - n_-^\mu) \,.
\label{addvecs}
\end{equation}
It follows that $v_+=e^{\vartheta/2}$ and $v_-=e^{-\vartheta/2}$, but
$v'_-=e^{\vartheta/2}$ and $v'_+=e^{-\vartheta/2}$. Similarly, in the
large-recoil limit ($\vartheta\gg 0$) the light quark in the initial
meson has large $p_+=\omega e^{\vartheta/2}$ and small $p_-$, whereas
the light quark in the final meson has large $p'_-=\omega'
e^{\vartheta/2}$ and small $p'_+$. Thus, the roles of the plus/minus
directions for the final meson are opposite to those for the initial
meson. The virtuality of the gluon and of the heavy quark are both
large: $k_g^2=(p'-p)^2\simeq -\omega\omega' e^\vartheta$, and $v\cdot
k_Q\simeq v\cdot p'\simeq \frac 12 \omega' e^\vartheta$
(Fig.~\ref{FIW}a) or $v'\cdot k_Q\simeq v'\cdot p\simeq \frac 12
\omega e^\vartheta$ (Fig.~\ref{FIW}b), respectively.

For large $\vartheta$, the contribution of the diagram in
Fig.~\ref{FIW}a to the matrix element in~(\ref{i3}), which defines
the Isgur--Wise function, is
\begin{eqnarray}
   2\pi\alpha_s\,\frac{C_F}{N_c}\,f^2\,e^{-2\vartheta}
   \int\frac{\text{d}\omega\,\text{d}\omega'}
   {\omega\omega^{\prime 2}}
   && \bar u(v') \left[ \varphi_+(\omega') \gamma_+
    e^{\vartheta/2} + \varphi_-(\omega') \gamma_- e^{-\vartheta/2} 
    \right] \rlap/v \nonumber\\
   &&\mbox{}\times \left[ \varphi_+(\omega) \gamma_- e^{\vartheta/2}
    + \varphi_-(\omega) \gamma_+ e^{-\vartheta/2} \right] u(v) \,,
\label{iw1}
\end{eqnarray}
where $\rlap/v=\frac 12\left(\gamma_- e^{\vartheta/2} + \gamma_+
e^{-\vartheta/2}\right)$. The factor $1/N_c$ arises from the
normalization of the colour wave functions of the meson states, which
is such that the matrix element in (\ref{i1}) is normalized to the
decay constant $f$. Using the relations in~(\ref{gampm}) together
with (\ref{addvecs}), we find that the leading contribution contains
the product $\varphi_+(\omega')\,\varphi_-(\omega)$, i.e.\ the
subleading-twist wave function is taken on the side where the gluon
exchange occurs. Adding the contribution of the diagram in
Fig.~\ref{FIW}b, we arrive at
\begin{equation}
   \xi(\cosh\vartheta) = 16\pi\alpha_s\,\frac{C_F}{N_c}\,f^2\,
   \langle\omega^{-2}\rangle_+\,\langle\omega^{-1}\rangle_-\,
   e^{-2\vartheta} \,.
\label{iw2}
\end{equation}

Based on our assumptions about the behaviour of the wave functions
for $\omega\to 0$ [cf.~(\ref{phiasy})], we expect that both
$\langle\omega^{-2}\rangle_+$ and $\langle\omega^{-1}\rangle_-$
diverge logarithmically at low $\omega$. This divergence is cut off
by the transverse momenta and virtualities of the light quarks in the
mesons, similar to the case of the $\pi$--$\rho$ form factor in
QCD~\cite{Chernyak}. This infrared sensitivity results in an
additional enhancement of the form factor, as can be seen by
replacing $q^2$ by $q^2-\Lambda^2$ in the gluon propagator (where
$\Lambda$ is of the order of a typical hadronic scale), which leads
to the replacement:
\begin{equation}
   \langle\omega^{-2}\rangle_+\,\langle\omega^{-1}\rangle_-
   \to \int \frac{\text{d}\omega\,\text{d}\omega'}
   {\omega\omega' + \Lambda^2 e^{-\vartheta}}\,
   \varphi'_+(\omega)\,\varphi_-(\omega')
   = \varphi'_+(0)\,\varphi_-(0)\,\frac{\vartheta^2}{2}
   + O(\vartheta) \,.
\label{iw3}
\end{equation}
Unfortunately, however, the subleading terms of order $\vartheta$
cannot be calculated without knowing details of the infrared cutoff.

Consider now the form factors $\xi_i$ and $\chi_i$ defined in
(\ref{i4a})--(\ref{i7}), which appear at order $1/m_Q$ in the
heavy-quark expansion. Since their contributions to the meson form
factors are suppressed by a power of $\Lambda/m_Q$ with respect to
the contribution of the Isgur--Wise function, they can only become
important if they have a slower fall-off at large $\vartheta$.
Therefore, it is sufficient to retain only those contributions where
the leading-twist wave function appears on both sides of the
diagrams. Then the meson helicity is conserved, and the gluon
polarization is orthogonal to the $v$--$v'$ plane. Let us first focus
on the functions $\xi_i$. The only way to get the leading-twist wave
function on both sides of the diagram is to attach the gluon to the
current operator, as shown by the diagram in Fig.~\ref{FIW}c.
Simplifying the resulting spinor product using (\ref{gampm}) and
(\ref{addvecs}), we obtain
\begin{equation}
   4\pi\alpha_s\,\frac{C_F}{N_c}\,f^2\,e^{-\vartheta}\,
   \bar u(v') \gamma^\mu u(v) 
   \int\frac{\text{d}\omega\,\text{d}\omega'}{\omega\omega'}\,
   \varphi_+(\omega)\,\varphi_+(\omega') \,,
\label{iw4}
\end{equation}
which means that only the function $\xi_3$ receives a leading-twist contribution. It is given by
\begin{equation}
   \xi_3(\cosh\vartheta) = 4\pi\alpha_s\,\frac{C_F}{N_c}\,f^2\,
   \langle\omega^{-1}\rangle_+^2\,e^{-\vartheta} \,.
\label{iw5}
\end{equation}
Here the integral is infrared convergent, as in the case of the pion
from factor in QCD
\cite{Excl1,Excl2,Excl3,Brodsky,Chernyak,Baier,Brev}. Next consider
the functions $\chi_i$. The conservation of the meson helicity
implies that there must be an odd number of $\gamma$ matrices in the
matrix element. Indeed, only the function $\chi_2$ in~(\ref{i7})
receives a leading contribution. Calculating the diagram in
Fig.~\ref{FIW}d, we find
\begin{equation}
   -8\pi\alpha_s\,\frac{C_F}{N_c}\,f^2\,e^{-2\vartheta}
   \int\frac{\text{d}\omega\,\text{d}\omega'}
   {\omega\omega^{\prime 2}}\,\varphi_+(\omega)\,
   \varphi_+(\omega')\,\bar u(v') (\gamma^\mu k_g^\nu
   - \gamma^\nu k_g^\mu) u(v) \,,
\label{iw6}
\end{equation}
where $k_g=\omega' v'-\omega v$ is the gluon momentum. Since, by
definition, the indices $\mu$ and $\nu$ are restricted to the
subspace orthogonal to $v$, only the first term in $k_g$ has to be
kept. Taking into account the definition of $\chi_2$ in
(\ref{chidef}) and (\ref{i7}), we obtain
\begin{equation}
   \chi_2(\cosh\vartheta) = - \xi_3(\cosh\vartheta)\,
   e^{-\vartheta} \,.
\label{iw7}
\end{equation}
Note that in the expressions for the meson form factors $\chi_2$ is
multiplied by $\cosh\vartheta$~\cite{review,Luke}, so that its
contributions are of the same order as the contributions of $\xi_3$.

We can compare our asymptotic results for the leading and subleading
Isgur--Wise functions in~(\ref{iw2}), (\ref{iw5}) and (\ref{iw7})
with the large-$\vartheta$ limit of the two-loop QCD sum-rule
expressions for these functions, which have been obtained
in~\cite{N2,NLN}. We find that the results of the sum-rule
calculations do indeed reproduce the correct asymptotic behaviour; in
particular, the relation~(\ref{iw7}) between $\chi_2$ and $\xi_3$ is
satisfied. Moreover, the sum rules allow us to determine the
normalization factors appearing in the expressions for $\xi$ in
(\ref{iw2}) and for $\xi_3$ in (\ref{iw5}). This is explained in
detail in Appendix~C. For later convenience, we also present the
expressions obtained using the model wave functions (\ref{sr26}).
They are:
\begin{equation}
   \xi(v\cdot v') \approx 3\pi\alpha_s\,\frac{f^2}{\bar\Lambda^3}\,
   \frac{\ln^2(v\cdot v')}{(v\cdot v')^2} \,, \quad
   \xi_3(v\cdot v') \approx 2\pi\alpha_s\,
   \frac{f^2}{\bar\Lambda^2}\,\frac{1}{v\cdot v'} \,.
\label{modfuns}
\end{equation}
Let us discuss the applicability regions for these asymptotic
results. QCD sum rules suggest that there are ``soft'' contributions
to the Isgur--Wise function which fall of like $1/(v\cdot v')^2$
\cite{Rad,N1,BS,BBG}. If this is correct, the asymptotic behaviour
given by (\ref{iw2}) and (\ref{iw3}) would dominate only if
$\ln(v\cdot v')\gg 1$ and $\alpha_s\ln^2(v\cdot v')\gg 1$. If the
``soft'' contributions vanish faster than $1/(v\cdot v')^2$, the
second requirement is removed. The fact that $\ln(v\cdot v')$ is, in
most practical applications, not a large parameter implies that the
asymptotic result for the Isgur--Wise function may be considered as a
rough estimate only. To reach the asymptotic regime would require
$v\cdot v'=O(100)$. QCD sum rules also suggest that the ``soft''
contributions to $\xi_3$ fall off like $1/(v\cdot v')^2$
\cite{N3,BaiG,NLN}, meaning that the leading hard contribution given
in (\ref{iw5}) is enhanced by a power of $v\cdot v'$. As a
consequence, our predictions for the functions $\xi_3$ and $\chi_2$
are much more accurate than for the Isgur--Wise function. The
asymptotic behaviour should set in when $\alpha_s\,v\cdot v'\gg 1$,
which requires $v\cdot v'=O(10)$.

An important aspect of physics is still missing from our discussion
of the asymptotic behaviour of meson form factors at large recoil.
Since the quarks receive a large acceleration during the transition
process, they emit gluon bremsstrahlung, which leads to an additional
damping of the transition amplitudes (Sudakov form factor). Because
the mesons are colourless, the double logarithms of the type
$[\alpha_s\ln^2(v\cdot v')]^n$ cancel in the expressions for the
meson form factors; however, single logarithms in $v\cdot v'$ remain,
which are enhanced by logarithms of the heavy-quark mass. They arise
from the emission of gluon bremsstrahlung with energies in the range
$\mu<E_g<m_Q$ ($\mu\ll m_Q$, see below). Thus, in perturbation theory
there are large double-logarithmic contributions of the type
$[\alpha_s\ln(m_Q/\mu)\ln(v\cdot v')]^n$ to the form factors. The
situation is similar to the case of the contributions to the pion
form factor coming from the region $x\to 0$, where almost all of the
pion momentum is carried by one quark \cite{CSZ}.

Because of the explicit dependence on the heavy-quark mass, these
large logarithms are not contained in the form factors of the HQET,
which are renormalized at a scale $\mu\ll m_Q$.\footnote{In practice,
the scale $\mu$ should be chosen such that there are no large
logarithms contained in the form factors of the HQET, but yet large
enough for perturbation theory to be valid. A typical choice is
$\mu\sim 1$~GeV.}
However, they appear when we relate the form factors of the HQET to
physical meson form factors using a perturbative matching procedure.
In this relation, there appear short-distance coefficient functions
$C_n(m_Q/\mu,v\cdot v')$, which can be calculated in
renormalization-group improved perturbation theory (see \cite{review}
for a review). In the case of transitions between two heavy hadrons,
the reparametrization invariance \cite{RPI} of the HQET ensures that
the coefficients multiplying the subleading functions $\chi_2$ and
$\xi_3$ are the same as the coefficients multiplying the
leading-order Isgur--Wise function $\xi$ \cite{RPIcur}. Indeed, in
leading logarithmic approximation (which is sufficient to control the
large logarithms mentioned above), all HQET form factors are
multiplied by a universal coefficient
\begin{equation}
   C(m_Q/\mu,v\cdot v') = \left( \frac{\alpha_s(m_Q)}{\alpha_s(\mu)}
   \right)^{a(v\cdot v')} \,,
\end{equation}
where ($v\cdot v'=\cosh\vartheta$) \cite{Falk,Poly,KoRa}
\begin{equation}
   a(v\cdot v') = \frac{2 C_F}{\beta_0} \left(
   \vartheta\coth\vartheta - 1 \right)
   = \frac{2 C_F}{\beta_0} \left[
   \vartheta - 1 + O(e^{-\vartheta}) \right] \,.
\end{equation}
For large recoil, we find that
\begin{equation}
   C\to \left( \frac{e}{2 v\cdot v'} \right)^\eta
   = e^{-\eta(\vartheta-1)} \,,
\end{equation}
where
\begin{equation}
   \eta = \frac{2 C_F}{\beta_0}
   \ln\frac{\alpha_s(\mu)}{\alpha_s(m_Q)} \,.
\label{etadef}
\end{equation}
This expression sums the large Sudakov logarithms correctly to all
orders in perturbation theory. The effect of this bremsstrahlung
correction is an additional power-like suppression of the physical
meson form factors for large values of $v\cdot v'$. Using $\mu\approx
1$~GeV, we find that for $B$-meson decays the power $\eta$ is given
by $\eta\approx 0.2$--0.3, i.e.\ the overall effect of bremsstrahlung
emission is rather small.

Using the results obtained in this section, it is straightforward to
derive the asymptotic behaviour of all form factors describing
current-induced transitions between any two pseudoscalar or vector
mesons containing a heavy quark. The relevant formulae, which relate
the meson form factors to the Isgur--Wise functions, can be found,
e.g., in~\cite{review,Luke,FN}. Here we restrict ourselves to the
results obtained for the matrix elements of the vector current
$V^\mu=\bar b\gamma^\mu b$ between $B$-meson states. We
find\footnote{To obtain the conventional relativistic normalization
of meson states, the right-hand sides in these equations have to be
multiplied by $m_{B^{(*)}}$.}
\begin{eqnarray}
   \langle B(v')|V^\mu|B(v)\rangle &=& h_+\,(v+v')^\mu \,,
    \nonumber\\
   \langle B^*(e',v')|V^\mu|B(v)\rangle &=& h_{\text{V}}\,
    \epsilon^{\mu\nu\alpha\beta} e^{\prime *}_\nu v'_\alpha
    v_\beta \,, \nonumber\\
   \langle B^*_{\text{L}}(v')|V^\mu|B^*_{\text{L}}(v)\rangle
   &=& h_{\text{L}}\,(v+v')^\mu \,, \nonumber\\
   \langle B^*_{\text{T}}(e',v')|V^\mu|B^*_{\text{T}}(e,v)\rangle
   &=& - h_{\text{T}}\,e\cdot e^{\prime *}\,
    (v+v')^\mu \,, \nonumber\\
   \langle B^*_{\text{T}}(e',v')|V^\mu|B^*_{\text{L}}(v)\rangle
   &=& \sinh\vartheta\,h_{\text{TL}}\,e^{\prime *\mu} \,,
\label{iw8}
\end{eqnarray}
where L and T refer to longitudinal (i.e.\ in the $v$--$v'$ plane)
and transverse (i.e.\ orthogonal to that plane) polarization states.
We find that, asymptotically,
\begin{eqnarray}
   h_+ &=& C \left( \xi - \frac{4}{m_b}\,\cosh\vartheta\,\chi_2
    \right) = C \left( \xi + \frac{2}{m_b}\,\xi_3 \right)
    \,, \nonumber\\
   h_{\text{V}} &=& C \left[ \xi - \frac{1}{m_b}\,
    (\xi_3 + 2\cosh\vartheta\,\chi_2) \right] = C \xi
    \,, \nonumber\\
   h_{\text{L}} &=& C \left( \xi + \frac{4}{m_b}\,\cosh\vartheta\,
    \chi_2 \right) = C \left( \xi - \frac{2}{m_b}\,\xi_3 \right)
    \,, \nonumber\\
   h_{\text{T}} &=& C \xi \,, \nonumber\\
   h_{\text{TL}} &=& C \left[ \xi + \frac{1}{m_b}\,
    (\xi_3 + 2\cosh\vartheta\,\chi_2) \right]  = C \xi \,.
\label{hiasy}
\end{eqnarray}
The most striking feature of these results is the fact that the form
factor for two longitudinally polarized $B^*$ mesons, which is
positive for $\cosh\vartheta\ll m_b/\Lambda$, becomes negative for
$\cosh\vartheta\gg m_b/\Lambda$, and hence has a zero at some
intermediate values $\cosh\vartheta\sim m_b/\Lambda$. Since for
spacelike (negative) values of $q^2$, corresponding to $v\cdot v'>1$,
all form factors are real, the existence of this zero is an exact
statement not affected by subleading corrections. We should stress
that this observation is not specific to heavy-light mesons. The form
factor of, say, longitudinally polarized $\rho$ mesons also has a
zero at some negative value of $q^2$ \cite{Chernyak}. For timelike
(positive) values of $q^2$, on the other hand, it is the form factor
of pseudoscalar $B$ mesons which has a zero, and this zero is
situated inside the physical region of the production of $B\bar B$
pairs in $e^+ e^-$ collisions. Strictly speaking, because form
factors at timelike values of $q^2$, corresponding to $v\cdot v'<-1$,
are complex, this zero is not absolutely exact. However, in our
approximation the imaginary part is negligible.\footnote{Note also
that any mechanism that leads to a common phase factor of the form
factors, such as final-state interactions, does not spoil the
existence of the zero.}

The model-independent results obtained in this section can be checked
in the simple model where a meson is built out of two heavy quarks
with masses $m$ and $\mu$ such that $m\gg\mu\gg\Lambda$. This is
discussed in detail in Appendix~D. The same model has been considered
by Brodsky and Ji~\cite{zero}, who observed for the first time the
zero of the pseudoscalar form factor in the physical region of large
positive $q^2$. However, their claim that the form factor of
longitudinally polarized vector mesons would have the same behaviour
is incorrect. Our analysis shows that this form factor has a zero in
the region of spacelike momentum transfer, i.e.\ for large negative
$q^2$.

\section{Applications}
\label{apps}

\begin{figure}
\epsfxsize=8cm
\centerline{\epsffile{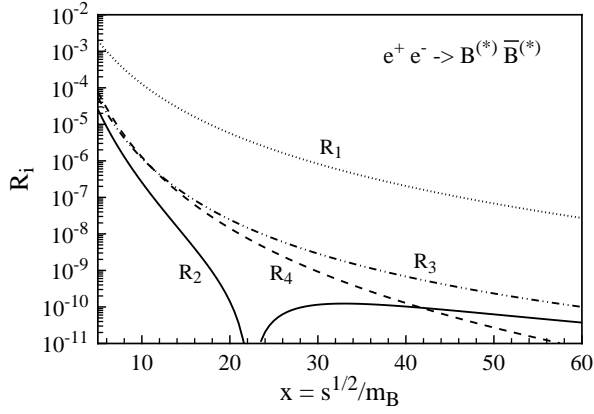}}
\vspace{0.2cm}
\caption{Cross-section ratios $R_i$ for $B\bar B$ pair production in
$e^+ e^-$ collisions at large $x$. We use the form factors in
(\protect\ref{ximodel}) with $\bar\Lambda=550$~MeV and $\alpha_s
f^2/\bar\Lambda^3=0.06$.}
\label{Rbratio}
\end{figure}

We finally apply our results to calculate the cross section for the
reaction $e^+ e^-\to B^{(*)}\bar B^{(*)}$ in the region $s\gg 4
m_B^2$. Recall from the introduction that the part of the
electromagnetic current that couples to light quarks does not give a
leading contribution in the asymptotic regime. Hence, it is justified
to use the relations in (\ref{hiasy}) for the relevant form factors.
As usual, we define
\begin{equation}
   R_X = \frac{\sigma(e^+ e^-\to X)}{\sigma(e^+ e^-\to\mu^+ \mu^-)}
   \,.
\label{iwx1}
\end{equation}
Using crossing symmetry and the matrix elements given in (\ref{iw8}),
we obtain for $x=\sqrt s/m_B\gg 1$:
\begin{eqnarray}
   R_1 &\equiv& R_{B B^* + B^*_{\text{T}} B^*_{\text{L}}}
    = \frac{z_b^2}{2}\,x^2 |C|^2 |\xi|^2 \,, \nonumber\\
   R_2 &\equiv& R_{B B} = \frac{z_b^2}{4}\,|C|^2
    \left| \xi + \frac{2}{m_b}\,\xi_3 \right|^2 \,, \nonumber\\
   R_3 &\equiv& R_{B^*_{\text{L}} B^*_{\text{L}}} = \frac{z_b^2}{4}\,
    |C|^2 \left| \xi - \frac{2}{m_b}\,\xi_3 \right|^2 \,, \nonumber\\
   R_4 &\equiv& R_{B^*_{\text{T}} B^*_{\text{T}}} = \frac{z_b^2}{2}\,
    |C|^2 |\xi|^2 \,,
\label{iw17}
\end{eqnarray}
where the form factors are functions of $v\cdot v'\simeq -\case 12
x^2$, and $|C|=(e/x^2)^{-\eta}$ with $\eta$ given in (\ref{etadef}).
Here $z_b=-\case 13$ is the electric charge of the $b$ quark. As an
illustration, we show in Fig.~\ref{Rbratio} our predictions for the
various $B^{(*)}\bar B^{(*)}$ production cross-sections as a function
of $x$ obtained using the model wave functions in (\ref{sr26}), for
which [cf.~(\ref{modfuns})]
\begin{equation}
   \xi(x) \approx 48\pi\,
   \frac{\alpha_s f^2}{\bar\Lambda^3}\,\frac{\ln^2\!x}{x^4} \,,
   \quad 
   \xi_3(x) \approx - 4\pi\,
   \frac{\alpha_s f^2}{\bar\Lambda^2}\,\frac{1}{x^2} \,.
\label{ximodel}
\end{equation}
For simplicity, we have neglected bremsstrahlung effects setting
$C=1$ in (\ref{iw17}). Because of the kinematic enhancement factor
$x^2$, the ratio $R_1$ generally dominates at large $x$. This means
that mostly $B B^*$ and $B^*_{\text{T}} B^*_{\text{L}}$ pairs are
produced, with $R\sim 1/s^3$ and angular distribution
$1+\cos^2\theta$ \cite{e+e-,FaGr}. More interesting from the point of
view of the present work are the three other ratios. The $B\bar B$
production cross section vanishes at some value
$x_0\sim\sqrt{m_b/\Lambda}$, since $\xi\sim\ln^2\!x/x^4$ and
$\xi_3\sim -\Lambda/x^2$. In our simple model, we find that:
\begin{equation}
   x_0\approx \sqrt{\frac{6 m_b}{\bar\Lambda}}\,\ln x_0
   \approx 25 \,.
\label{iw18}
\end{equation}
We stress again that the accuracy of this prediction is not high,
because of our poor knowledge of the asymptotic behaviour of the
Isgur--Wise function. For $1\ll x\ll x_0$, the ratios $R_2$, $R_3$
and $R_4$ are all of the same magnitude and scale like $1/x^8$. This
situation had been studied previously in the context of the
HQET~\cite{e+e-,FaGr}. For $x\gg x_0$, however, another pattern sets
in. Then the contribution of $\xi_3$ to the ratios $R_2$ and $R_3$
dominates over the contribution from the Isgur--Wise function $\xi$,
so that $R_2$ and $R_3$ scale like $1/x^4$ and dominate over $R_4$.
In principle, at very large $x$ the ratios $R_2$ and $R_3$ should
even dominate over $R_1$, which scales like $1/x^6$. However, because
of the double-logarithmic enhancement of the Isgur--Wise function
this would require enormous values $x>x_1$, where in our model $x_1$
is given by
\begin{equation}
   x_1\approx \frac{6 m_b}{\bar\Lambda}\,\ln^2\!x_1 \approx 3500 \,.
\end{equation}
In the ultra-asymptotic region $x\gg x_1$, mostly $B B$ and
$B^*_{\text{L}} B^*_{\text{L}}$ pairs would be produced, with $R\sim
1/s^2$ and angular distribution $\sin^2\vartheta$.

\begin{figure}
\epsfxsize=8cm
\centerline{\epsffile{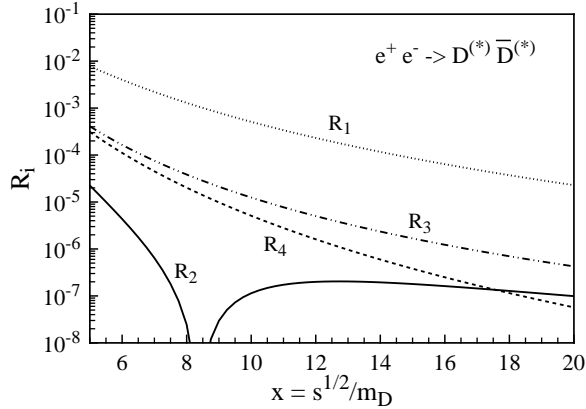}}
\vspace{0.2cm}
\caption{Cross-section ratios $R_i$ for $D\bar D$ pair production in
$e^+ e^-$ collisions at large $x$.}
\label{Rcratio}
\end{figure}

These qualitative features, which are independent of the particular
choice adopted for the meson wave functions, are clearly exhibited in
Fig.~\ref{Rbratio}. Unfortunately, however, the cross sections for
$B^{(*)}\bar B^{(*)}$ production at large $x$ are so small that they
will most likely be irrelevant to experiments. The situation is
somewhat more favourable in the case of the pair production of charm
mesons. We can apply our results to this case by performing obvious
substitutions ($m_b\to m_c$, $z_b\to z_c=\case 23$, etc.) in the
above formulae. We then find that $x_0\approx 8$ in the case of charm
pair production, corresponding to moderate energies of order 15~GeV.
The second turnover point is, however, still too high ($x_1\approx
650$) to be of any interest. The resulting cross sections are shown
in Fig.~\ref{Rcratio}.

In summary, we have applied methods developed for hard exclusive QCD
processes to calculate the asymptotic behaviour of heavy-meson form
factors at large recoil. We find that this behaviour is determined by
the leading- and subleading-twist meson wave functions. For $1\ll
|v\cdot v'|\ll m_Q/\Lambda$, the form factors are dominated by the
Isgur--Wise function $\xi$, which is determined by the interference
between the wave functions of leading and subleading twist. At
$|v\cdot v'|\gg m_Q/\Lambda$, they are dominated by the two functions
$\xi_3$ and $\chi_2$ arising at order $1/m_Q$ in the heavy-quark
expansion, which are determined by the leading-twist wave function
alone. The sum of these contributions describes the form factors in
the whole region $|v\cdot v'|\gg 1$. Central objects of our study are
the meson wave functions $\varphi_\pm(\omega)$, which are defined in
terms of the Fourier transforms of the matrix element of bilocal
operators on the light-cone. We have derived the (Brodsky--Lepage)
evolution equations obeyed by these wave functions, and we have
investigated the properties of the wave functions (such as their
moments) using QCD sum rules. Finally, we have discussed as an
application the implications of our results for the production of
heavy-meson pairs in $e^+ e^-$ collisions.

\acknowledgments
We are grateful to G.P.~Korchemsky, C.T.~Sachrajda, and T.~Mannel for
useful comments. A.G.G.\ is deeply indebted to S.V.~Mikhailov for
detailed discussions, in the course of which some ideas of this work
were produced, and to V.L.~Chernyak for thorough discussions. He also
likes to acknowledge the hospitality of the CERN Theory Division,
where the work on this paper was completed. M.N. is grateful to
V.N.~Gribov for enlightening discussions about gluon bremsstrahlung
in heavy-quark transitions.

\setcounter{equation}{0}
\renewcommand{\theequation}{A\arabic{equation}}
\section*{Appendix A: Covariant Trace Formalism}

The most convenient way to calculate the matrix elements of operators
between the physical pseudoscalar and vector $(q\bar Q)$ meson states
(rather than the spin-$\case 12$ mesons obtained when the heavy-quark
spin is switched off) is provided by the covariant tensor formalism
introduced in~\cite{Falk}. In the HQET, the spin wave function of the
ground-state meson doublet is described by the $4\times 4$ Dirac
matrix
\begin{equation}
   {\cal M}(v) = \frac{1+\rlap/v}{2}\left\{ \begin{array}{ll}
   -i\gamma_5 & \mbox{; pseudoscalar meson $M(v)$,} \\
   \rlap/e & \mbox{; vector meson $M^*(e,v)$,} \\
   \end{array} \right.
\end{equation}
where $v$ is the meson velocity, and $e$ is the polarization vector
of the vector meson ($e\cdot v=0$). The matrix ${\cal M}(v)$ simply
contains the appropriate spin-parity projections of the spinor
product $u_q(v)\,\bar v_Q(v)$~\cite{AFal}. It satisfies
\begin{equation}
   \rlap/v\,{\cal M}(v) = {\cal M}(v){\cal M}(v)
   = - {\cal M}(v)\,\rlap/v \,.
\end{equation}

Operator matrix elements between meson states can be represented by
traces over these wave functions. Consider first the matrix elements
of heavy-light operators of the type $\bar Q_v\Gamma\,{\bf
O}(iD)\,q$, where $\Gamma$ is an arbitrary Dirac matrix, and ${\bf
O}(iD)$ is a differential operator acting on the light-quark field,
between a meson state and the vacuum. Their representation is
\begin{equation}
   \langle\,0\,|\bar Q_v\Gamma\,{\bf O}(iD)\,q|M(v)\rangle
   = \text{Tr}\{O(v)\,{\cal M}(v)\,\Gamma \} \,,
\end{equation}
where $O(v)$ is the most general matrix with the same transformation
properties (under the Lorentz group and heavy-quark symmetry) as the
operator ${\bf O}$. In the spinor formalism adopted in our paper,
the same matrix element would read
\begin{equation}
   \langle\,0\,|Q_v^*\,{\bf O}(iD)\,q|M(v)\rangle = O(v)\,u(v) 
\end{equation}
with the same matrix $O(v)$. The covariant decomposition of this
matrix determines the number of reduced matrix elements (generalized
Isgur--Wise form factors) that appear in the heavy-quark expansion.
As an example, we give the expressions in the trace formalism which
correspond to the definitions in~(\ref{i1}), (\ref{Dmumat}), and
(\ref{theta}):
\begin{eqnarray}
   \langle\,0\,|\bar Q_v\Gamma\,q|M(v)\rangle
   &=& f\,\text{Tr}\{{\cal M}(v)\,\Gamma \} \,, \nonumber\\
   \langle\,0\,|\bar Q_v\Gamma\,iD^\mu q|M(v)\rangle
   &=& \case 13 f\bar\Lambda\,\text{Tr}\{ (4 v^\mu-\gamma^\mu)\,
    {\cal M}(v)\,\Gamma \} \,, \nonumber\\
   \langle\,0\,|\bar Q_v\Gamma\,iD^\mu iD^\nu q|M(v)\rangle
   &=& f\,\text{Tr}\{ \Theta^{\mu\nu}(v)\,{\cal M}(v)\,\Gamma
    \} \,,
\end{eqnarray}
where $\Theta^{\mu\nu}(v)$ is given in (\ref{Thmunu}).

In the end of Sec.~\ref{WF}, we need the generalization of
(\ref{wf2}) in the trace formalism. It reads:
\begin{equation}
   \langle\,0\,|\bar Q_v(0)\Gamma\,E(0,z)\,q(z)|M(v)\rangle
   = f\,\text{Tr}\Big\{ \Big[ \widetilde\varphi_+(t)
   + \frac{1}{2t}\,[\widetilde\varphi_-(t)
   - \widetilde\varphi_+(t)]\,\rlap/z \Big]\,
   {\cal M}(v)\,\Gamma \Big\} \,.
\end{equation}
Evaluating the trace for various choices of $\Gamma$, we recover the
results given in (\ref{phidef1})--(\ref{norm2}).

The trace formalism is readily extended to more complicated cases,
such as transition matrix elements between two meson states. For
instance, the expressions corresponding to the definitions
in~(\ref{i3}) and (\ref{i4a}) read:
\begin{eqnarray}
   \langle M(v')|\bar Q_v\Gamma\,Q_{v'}|M(v)\rangle
   &=& \xi(v\cdot v')\,\text{Tr}\{ {\cal M}(v)\,\Gamma\,
    {\overline{\cal M}}(v') \} \,, \nonumber\\
   \langle M(v')|(i D^{\mu\dagger} \bar Q_v)\Gamma\,Q_{v'}
   |M(v)\rangle &=& \text{Tr}\{ \xi^\mu(v,v')\,{\cal M}(v)\,
    \Gamma\,{\overline{\cal M}}(v') \} \,,
\end{eqnarray}
where the covariant decomposition of $\xi^\mu(v,v')$ is given in
(\ref{i4}). Note that as a consequence of the fact that in this paper
we work with $(q\bar Q)$ rather than $(Q\bar q)$ mesons, the trace
formalism is slightly different from the one usual employed in the
literature~\cite{review,Luke,FN}. Crossing symmetry implies that the
form factors for $(q\bar Q)$ mesons are related to those for $(Q\bar
q)$ mesons by Hermitean conjugation followed by the substitutions
$v\to -v$ and $v'\to -v'$. We have defined the invariant functions in
the HQET ($\xi$, $\xi_3$, $\chi_i$ etc.) in such a way that the
resulting expressions for the physical matrix elements look the same
as in the conventional formalism.

\setcounter{equation}{0}
\renewcommand{\theequation}{B\arabic{equation}}
\section*{Appendix B: Non-local Condensates}

The contributions of higher-order non-perturbative corrections to the
sum rules for $\lambda_E^2$ and $\lambda_H^2$ in (\ref{sr9}) can be
included by introducing two functions, $f^{(1)}(x^2)$ and
$f^{(2)}(x^2)$, which parametrize the following non-local
condensates~\cite{Higher}:
\begin{eqnarray}
   f^{(1)}(x^2) &=& \frac{\langle\bar q(0) E(0,x) \sigma_{\mu\nu} 
     G^{\mu\nu}(x) q(x)\rangle}
    {\langle\bar q\,\sigma_{\mu\nu} G^{\mu\nu} q\rangle}
    = 1 + \frac{Q_1 - Q_2 - 2 Q_3}{m_0^2\langle\bar q q\rangle}\,
    \frac{x^2}{8} + \dots \,, \nonumber\\
   f^{(2)}(x^2) &=& \frac{4 x^\alpha x_\beta}{x^2}\,
    \frac{\langle\bar q(0) E(0,x) \sigma_{\mu\alpha} 
     G^{\mu\beta}(x) q(x)\rangle}
    {\langle\bar q\,\sigma_{\mu\nu} G^{\mu\nu} q\rangle} 
    = 1 + \frac{2 Q_1 - Q_2 - 3 Q_3}{m_0^2\langle\bar q q\rangle}\,
    \frac{x^2}{12} + \dots \,.
\label{sr7}
\end{eqnarray}
The quantities $Q_i$ form a basis of dimension-7 quark--gluon
condensates and are defined as ($\tilde G^{\mu\nu}=\frac 12
\epsilon^{\mu\nu\alpha\beta} G_{\alpha\beta}$):
\begin{eqnarray}
   Q_1 &=& \langle\bar q G_{\mu\nu} G^{\mu\nu} q\rangle \,,
    \nonumber\\
   Q_2 &=& i\langle\bar q G_{\mu\nu} \tilde G^{\mu\nu}
    \gamma_5 q\rangle \,, \nonumber\\
   Q_3 &=& i\langle\bar q\sigma_{\mu\nu} G^{\mu\lambda}
    G^\nu{}_\lambda q\rangle \,, \nonumber\\
   Q_4 &=& \langle\bar q\sigma_{\mu\nu} (D^\mu D_\alpha
    G^{\nu\alpha}) q\rangle \,.
\end{eqnarray}
If these corrections are included, the sum rules (\ref{sr9}) are
modified in the following way:
\begin{eqnarray}
   f^2\lambda_E^2 e^{-\bar\Lambda\tau} &=& - N_c C_F\,
    \frac{\alpha_s}{\pi^3\tau^5}\,\delta_4(\varepsilon_c\tau)
    - \frac{m_0^2\langle\bar q q\rangle}{16}\,f^{(2)}(-\tau^2)
    \,, \nonumber\\
   f^2\lambda_H^2 e^{-\bar\Lambda\tau} &=& - N_c C_F\, 
    \frac{\alpha_s}{2\pi^3\tau^5}\,\delta_4(\varepsilon_c\tau)
    - C_F\,\frac{3\alpha_s}{4\pi\tau^2}\,
    \langle\bar q q\rangle\,\delta_1(\varepsilon_c\tau)
    \nonumber\\
   &&\mbox{}+ \frac{\alpha_s\langle G^2\rangle}{16\pi\tau}\,
    \delta_0(\varepsilon_c\tau) 
    - \frac{m_0^2\langle\bar q q\rangle}{16}\,\left[
    2 f^{(1)}(-\tau^2) - f^{(2)}(-\tau^2) \right] \,.
\end{eqnarray}

\widetext
Next we give some details of the calculation of the quark-condensate
contributions to the QCD sum rules for the meson wave functions. The
contribution with the cut light-quark line shown in Fig.~\ref{Fsr2}c
involves a trilocal object, the non-collinear quark
condensate~\cite{Higher}:
\begin{eqnarray}
   &&\langle\bar q_\beta(y) E(y,0) E(0,x) q_\alpha(x)\rangle
    \nonumber\\
   &&\quad = \frac{\langle\bar q q\rangle}{4}\,\left[
    f_S(x,y) + \frac{m_0^2}{48}\,[\rlap/x,\rlap/y]\,f_T(x,y)
    - \frac{i}{4}[ \rlap/x\, f_V(x,y) - \rlap/y\,f_V(y,x) ]
    \right]_{\alpha\beta} \,.
\label{sr14}
\end{eqnarray}
Neglecting the function $f_V(x,y)$, whose operator product expansion
contains even-dimen\-sional quark condensates with $d\ge 6$ and whose
contribution to heavy-meson sum rules is
negligible~\cite{N1,Patricia,David,Shuryak}, we obtain
\begin{equation}
   \Pi_\pm^{(2)}(v\cdot x,t) = -\case 14\,
   \langle\bar q q\rangle
   \left[ f_S(z,-x) \pm \case{1}{24}\,m_0^2 v\cdot x\,t\,
   f_T(z,-x) \right] \,,
\label{sr15}
\end{equation}
where $z^2=0$, $x^2=(v\cdot x)^2$, and $z\cdot x=v\cdot x\,t$.

The non-collinear condensate in~(\ref{sr14}) can be expanded in $x$
at fixed $y$. One finds~\cite{Higher}:
\begin{eqnarray}
   f_S(x,y) &=& f_S[(y-x)^2] + \frac 43\,
    \frac{(xy)^2-x^2 y^2}{y^2} \left[ f_S'(y^2)
    + \frac{y^2}{2}\,f_S''(y^2) - \frac{m_0^2}{16}\,f^{(1)}(y^2)
    \right] + O(x^3) \,, \nonumber\\
   f_T(x,y) &=& \frac{16}{m_0^2} \left\{ f_S'(y^2)
    - \frac{xy}{y^2} \left[ f_S'(y^2) + 2 y^2\,f_S''(y^2)
    - \frac{m_0^2}{16}\,f^{(2)}(y^2) \right] \right\} + O(x^2) \,,
\label{sr16}
\end{eqnarray}
where the function $f_S(x^2)$ parametrizes the bilocal quark
condensate and is given by~\cite{Mikh,Higher}
\begin{equation}
   f_S(x^2) = \frac{\langle\bar q(0) E(0,x) q(x)\rangle}
   {\langle\bar q q\rangle} = 1 + \frac{m_0^2 x^2}{16}
   + \frac{6 Q_1 - 3 Q_2 - 6 Q_3 + 2 Q_4}{\langle\bar q q\rangle}\,
   \frac{x^4}{1152} + O(x^6) \,.
\label{sr17}
\end{equation}
\narrowtext\noindent
A convenient representation of the bilocal quark condensate
is~\cite{Mikh}
\begin{equation}
   f_S(x^2) = \int\text{d}\nu\,\widetilde f_S(\nu)\,e^{\nu x^2} \,,
\label{sr20}
\end{equation}
where
\begin{equation}
   \int\text{d}\nu\,\widetilde f_S(\nu) = 1 \,,\quad
   \int\text{d}\nu\,\nu\,\widetilde f_S(\nu) = \frac{m_0^2}{16} \,,
\label{sr20b}
\end{equation}
and so on. The function $\widetilde f_S(\nu)$ can be interpreted as
the distribution quarks with virtuality $\nu$ in the QCD vacuum. The
local expansion in~(\ref{sr17}) corresponds to the expansion
\begin{equation}
   \widetilde f_S(\nu) = \delta(\nu) - \frac{m_0^2}{16}\,
   \delta'(\nu) + \dots \,.
\label{sr21}
\end{equation}
Because the factor $E(0,x)$ in~(\ref{sr17}) can be interpreted as the
heavy-quark propagator, the asymptotic behaviour of $f_S(x^2)$ at
large $-x^2$ is
\begin{equation}
   f_S(x^2) \sim e^{-\bar\Lambda\sqrt{-x^2}} \,.
\label{sr22}
\end{equation}
This fixes the behaviour of $\widetilde f_S(\nu)$ for $\nu\to 0$. A
simple ansatz for the distribution function, which satisfies this
constraint, was proposed in~\cite{Bakulev}:
\begin{equation}
   \widetilde f_S(\nu) = N \exp\left( -\frac{\bar\Lambda^2}{4\nu}
   - \sigma\nu \right) \,,
\label{sr23}
\end{equation}
where $N$ and $\sigma$ are fixed by the conditions~(\ref{sr20b}).

\setcounter{equation}{0}
\renewcommand{\theequation}{C\arabic{equation}}
\section*{Appendix C: Form-Factor Asymptotics from QCD Sum Rules}

It is instructive to compare our asymptotic results for the leading
and subleading Isgur--Wise functions with the large-$\vartheta$ limit
of the two-loop (order-$\alpha_s$) QCD sum-rule expressions for these
functions, which have been obtained in~\cite{N2,NLN}. For very large
recoil, the three-point correlators considered in these sum rules
factorize into the convolution of the two-point
correlators~(\ref{sr18}) with hard-scattering amplitudes. In this
limit, only diagrams with a gluon exchange between a heavy quark and
the light quark remain. We find that the results of the sum-rule
calculations do indeed reproduce the correct asymptotic behaviour; in
particular, the relation~(\ref{iw7}) between $\chi_2$ and $\xi_3$ is
satisfied. For the normalization factors appearing in the expressions
for $\xi$ in (\ref{iw2}) and for $\xi_3$ in (\ref{iw5}), we obtain
from \cite{N2,NLN}:
\begin{eqnarray}
   f^2 \varphi'_+(0)\,\varphi_-(0)
   &=& \frac{N_c^2}{4\pi^4 f^2\tau^3}\,e^{\bar\Lambda\tau}\, 
    \delta_2(\varepsilon_c\tau) \,, \nonumber\\
   f^2 \langle\omega^{-1}\rangle_+^2
   &=& \frac{N_c^2}{\pi^4 f^2\tau^4}\,e^{\bar\Lambda\tau}\,
    \delta_3(\varepsilon_c\tau) \,.
\label{iw12}
\end{eqnarray}
We have retained the leading perturbative contributions only, since
the relevant non-local condensates have not yet been calculated to
order $\alpha_s$. Note, in particular, that the leading
quark-condensate contribution to the sum rule for the Isgur--Wise
function is constant and seems to dominate for large recoil. However,
once the non-locality of the quark condensate is taken into account,
one finds that this contribution actually vanishes quickly at large
recoil \cite{Rad,N1}.

\begin{figure}
\epsfxsize=5cm
\centerline{\epsffile{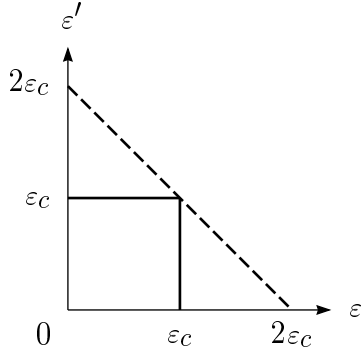}}
\caption{Square model (solid line) and triangle model (dashed line)
for the continuum subtraction in QCD sum rules for three-point
correlation functions.}
\label{conti}
\end{figure}

It is straightforward reproduce the expressions in (\ref{iw12})
starting from the sum-rule results for the wave functions
$\varphi_\pm(\omega)$ obtained in Sec.~\ref{SR}. The sum rule for the
product $\varphi_+(\omega)\varphi_\pm(\omega')$ at equal Borel
parameters has the form
\begin{equation}
   f^4 \varphi_+(\omega)\,\varphi_\pm(\omega')\,
   e^{-\bar\Lambda\tau}
   = \int\text{d}\varepsilon\,\text{d}\varepsilon'\,
   \rho_+(\omega,\varepsilon)\,\rho_\pm(\omega',\varepsilon')\,
   e^{-(\varepsilon+\varepsilon')\tau/2} \,,
\label{C2}
\end{equation}
where $\rho_\pm(\omega,\varepsilon)$ are the spectral densities of
the correlators (\ref{sr18}), and the integral is taken over the
complement of the continuum region. The precise form of the result
will depend on the particular way in which the continuum subtraction
is performed. The spectral densities for three-point correlators
depend on two variables, $\varepsilon$ and $\varepsilon'$; see
Fig.~\ref{conti}. The ``square model'' of the continuum subtraction
amounts to cutting off the integrals over these variables at the
threshold $\varepsilon_c$. This leads to an exact factorization of
the integrals in (\ref{C2}). Using the appropriate products of the
sum rules for the wave functions given in (\ref{sr25}), and retaining
the leading perturbative contributions only, we then obtain:
\begin{eqnarray}
   f^2 \varphi'_+(0)\,\varphi_-(0)
   &=& \frac{N_c^2}{4\pi^4 f^2\tau^3}\,e^{\bar\Lambda\tau}\, 
    \delta_0(\case 12\varepsilon_c\tau)\,
    \delta_1(\case 12\varepsilon_c\tau) \,, \nonumber\\
   f^2 \langle\omega^{-1}\rangle_+^2
   &=& \frac{N_c^2}{\pi^4 f^2\tau^4}\,e^{\bar\Lambda\tau}\,
    \Big[ \delta_1(\case 12\varepsilon_c\tau) \Big]^2 \,.
\label{iw12b}
\end{eqnarray}
On the other hand, it is well known that for small recoil the square
model of the continuum subtraction is inconsistent, as it leads to an
unphysical infinite slope of the Isgur--Wise function at $v\cdot
v'=1$ \cite{N1}. This deficiency is removed by using the ``triangle
model'', where $0<\varepsilon+\varepsilon'<2\varepsilon_c$, while the
difference $\varepsilon-\varepsilon'$ is unconstrained \cite{N1,BS}.
The triangle model was adopted in the calculations in \cite{N2,NLN}.
If we use it to evaluate (\ref{C2}), we indeed recover (\ref{iw12}).
This is a strong check of both, the present approach and the two-loop
calculations performed in \cite{N2,NLN}. However, in this case the
resulting sum rule for the product of the wave functions is no longer
exactly factorizable, meaning that the triangle model is not fully
consistent at large recoil, and so the square model is preferrable.
Comparing (\ref{iw12}) with (\ref{iw12b}), we observe that both
results agree in the limit $\varepsilon_c\to\infty$, when the choice
of the continuum model becomes irrelevant ($\delta_n\to 1$).

The sum rule for the Isgur-Wise function \cite{N1} has a built-in
model of the infrared cutoff, and therefore it allows us to estimate
the subleading $O(\vartheta)$ term to the Isgur--Wise function in
(\ref{iw3}). The result is
\begin{equation}
   \xi(\cosh\vartheta) \sim e^{-2\vartheta}\,\Big\{ \vartheta^2
   + \left[ 4 L(\varepsilon_c\tau)-5 \right]\,\vartheta
   + \dots \Big\} \,,
\end{equation}
where
\begin{equation}
   L(x_c) = \frac{\int\limits_0^{x_c}\!\text{d}x\,x^2 e^{-x}
    \ln 2x}{\int\limits_0^{x_c}\!\text{d}x\,x^2 e^{-x}} \,.
\end{equation}
In the relevant region of values of $\varepsilon_c\tau\approx 2.5$,
we find that $(4L-5)\approx -0.7$, meaning that the subleading term
is negative and has a coefficient of order unity.

\setcounter{equation}{0}
\renewcommand{\theequation}{D\arabic{equation}}
\section*{Appendix D: Static Quark Model}

Our model-independent results in~(\ref{iw2}), (\ref{iw5}),
(\ref{iw7}), and (\ref{iw8}) can be checked in the simple model where
a meson is composed of two heavy quarks with masses $m$ and $\mu$,
such that $m\gg\mu\gg\Lambda$~\cite{Seoul}. In this case,
$\varphi_+(\omega)=\varphi_-(\omega)=\delta(\omega-\mu)$, and
$\langle\omega^n\rangle_\pm=\mu^n$. Note that the
formulae~(\ref{wf9}) and (\ref{wf12}) for the lowest moments are
based on the equation of motion for a massless quark, and are thus no
longer applicable.

Let us consider the pseudoscalar form factor $h_+$ in (\ref{iw8}). It
is convenient to calculate it from the relation
$(\cosh\vartheta+1)\,h_+=\langle B(v')|v_\mu V^\mu|B(v)\rangle$. A
simple evaluation of the diagrams in Fig.~\ref{FIW}a,b gives
\begin{equation}
   h_+ = 2\pi\alpha_s\,\frac{C_F}{N_c}\,\frac{f^2}{\mu^2}\,
   e^{-2\vartheta}\,\text{Tr}\left[ \gamma_\mu\,\Phi\,\gamma^\mu
   S\,\rlap/v\,\bar\Phi' + \Phi\,\rlap/v\,S'\gamma_\mu\bar\Phi'
   \gamma^\mu \right] \,,
\label{iw9}
\end{equation}
where $\Phi=\gamma_5(1-\rlap/v)$ and
$\bar\Phi'=-(1-\rlap/v')\gamma_5$ are the spin structures arising for
pseudoscalar mesons. The heavy-quark propagators in Fig.~\ref{FIW}a,b
are given by
\begin{equation}
   S = \frac{m(1-\rlap/v) + \mu\rlap/v'}{-m\mu e^\varphi}
\label{iw10}
\end{equation}
and $S'=S(v\leftrightarrow v')$. If we retain the leading term
proportional to $m$ in the numerator of~(\ref{iw10}), then the gluon
is longitudinally polarized; the two diagrams contribute equally, and
we recover the contribution of the function $\xi$ in (\ref{iw2}) to
the form factor $h_+$ in (\ref{hiasy}). If, on the other hand, we
retain the term with $\rlap/v'$, we lose a factor $\mu/m$ but gain a
factor $e^\vartheta$ from the trace. Then only the first diagram
contributes; the gluon is transversely polarized, and we recover the
contribution of the function $\xi_3$ in (\ref{iw5}) to the form
factor $h_+$. In this model, the form factor of pseudoscalar mesons
has a zero at $q^2=2m^3/\mu$. The calculation can be repeated for
vector mesons using the spin structures $\Phi=\rlap/e(1-\rlap/v)$ and
$\bar\Phi'=(1-\rlap/v')\rlap/e'$. In particular, we find that the
form factor of longitudinally polarized vector mesons has a zero at
$q^2=-2m^3/\mu$. Other form factors can be calculated in a similar
way; our results agree with~\cite{Seoul}.

\end{document}